\documentclass[useAMS,usenatbib]{mn2e}
\usepackage{amssymb}
\usepackage{amsmath}
\usepackage{times}
\usepackage{graphicx}
\newcommand{\mincir}{\raise
-2.truept\hbox{\rlap{\hbox{$\sim$}}\raise5.truept 
\hbox{$<$}\ }}
\newcommand{\magcir}{\raise
-2.truept\hbox{\rlap{\hbox{$\sim$}}\raise5.truept
\hbox{$>$}\ }}
\newcommand{\minmag}{\raise-2.truept\hbox{\rlap{\hbox{$<$}}\raise
6.truept\hbox
{$>$}\ }}

\newcommand{\lya}{Lyman-$\alpha$~}
\newcommand{\gad} {{\small {GADGET-2}}\,}

\newcommand{\be}{\begin{equation}}
\newcommand{\ee}{\end{equation}}
\newcommand{\ba}{\begin{eqnarray}}
\newcommand{\ea}{\end{eqnarray}}
\newcommand{\brr}{\begin{array}}
 
\newcommand{\err}{\end{array}}
\newcommand{\bc}{\begin{center}}
\newcommand{\ec}{\end{center}}

\newcommand{\msun}{$M_{\odot}$}
\newcommand{\zsun}{$Z_{\odot}$}

\DeclareMathAlphabet{\mathsc}{OT1}{cmr}{m}{sc}
\def\testbx{bx}%
\DeclareRobustCommand{\ion}[2]{%
\relax\ifmmode
\ifx\testbx\f@series
{\mathbf{#1\,\mathsc{#2}}}\else
{\mathrm{#1\,\mathsc{#2}}}\fi
\else\textup{#1\,{\mdseries\textsc{#2}}}%
\fi}

\title[Damped \lya systems in high-resolution hydrodynamical
  simulations]{Damped \lya systems in high-resolution hydrodynamical
  simulations}

\author[E. Tescari, M. Viel, L. Tornatore \&
  S. Borgani]{E. Tescari$^{1,2,3}$, M. Viel$^{1,2}$, L. Tornatore$^{1,2,3}$
  \& S. Borgani$^{1,2,3}$ \\ $^1$ INAF - Osservatorio Astronomico di
  Trieste, Via G.B. Tiepolo 11, I-34131 Trieste, Italy \\ $^2$
  INFN/National Institute for Nuclear Physics, Via Valerio 2, I-34127
  Trieste, Italy \\ $^3$ Dipartimento di Astronomia dell'Universit\`a
  di Trieste, Via G.B. Tiepolo 11, I-34131 Trieste, Italy}

\begin{document}

\maketitle

\begin{abstract}
We investigate the properties of Damped \lya systems (DLAs) using
high-resolution and large box-size cosmological hydrodynamical
simulations of a $\Lambda$CDM model. The numerical code used is a
modification of \gad with a self consistent implementation of the
metal enrichment mechanism (Tornatore et al. 2007).  We explore the
numerical convergence of some relevant physical quantities and we vary
the parameters describing the properties of galactic winds; the
initial stellar mass function; the linear dark matter power spectrum
and the metal enrichment pattern of the IGM (Intergalactic Medium)
around DLAs. We focus on the properties of dark matter haloes that are
likely to be the hosts of DLAs systems: we predict relatively low star
formation rates ($\sim 0.01-0.1$ \msun /year) and metallicities around
0.1 \zsun, at least for the bulk of our haloes of masses between
$10^9$ and $10^{10}$ $h^{-1}M_{\rm \odot}$ hosting DLAs. For more
massive haloes metallicities and star formation rates depend on the
specific wind model.  We find that strong galactic winds with speed of
about 600 km/s, in an energy-driven wind scenario, are needed in order
to match the observed column density distribution function for DLAs
and the evolution of the neutral hydrogen content with redshift. The
momentum-driven implementation of the galactic wind model, that
relates the speed and mass load in the wind to the properties of the
dark matter haloes, shows a behaviour which is intermediate between
the energy-driven galactic winds of small ($\sim 100$ km/s) and large
($\sim 600$ km/s) velocities.  At $z=3$ the contribution of haloes of
masses between $10^9$ and $10^{10}$ $h^{-1}M_{\rm \odot}$, for DLAs
below $10^{20.8}$ cm$^{-2}$, to the column density distribution
function, is significant. By interpolating physical quantities along
line-of-sights through massive haloes we qualitatively show how
different galactic wind models impact on the IGM around DLAs.
Furthermore, we analyse statistics related to the velocity widths of
SiII associated to DLAs: while the expanding shells of gaseous matter
associated to the wind can account for the observed velocities, the
metallicity in the wind seems to be rather clumpy and this produces an
underestimation of the observed velocity widths.  We outline possible
solutions to this problem.
\end{abstract}


\section{Introduction}
Damped \lya systems (DLAs) are defined as quasar (QSO) absorption
systems with neutral hydrogen column density $N_{\rm HI} > 2\times
10^{20}$ cm$^{-2}$ \citep{wolfe86}. DLAs are considered as an
important reservoir and/or sink of gas for the galaxy formation
process in the high redshift universe and their neutral hydrogen
content dominate the total neutral hydrogen budget over a large
fraction of the cosmic history. The interplay between DLAs and
galaxies is thereby fundamental and should be addressed by any galaxy
formation model.

Significant observational efforts have been made in order to
understand the nature of DLAs
\citep[e.g.][]{wolfe95,storrie-lombardi00,raoturnshek00,prochaska01,peroux03,chenlanzetta03,prochaska05},
while, on the theoretical side, semi-analytical models and
high-resolution hydrodynamical simulations are routinely performed to
investigate the relation between DLAs and dark matter haloes and to
match their observed properties
\citep[e.g.][]{katz96,gardner97,haehnelt98,maller01,gardner01,nagamine04,okoshi05,nagamine07,pontzen08,barnes08}.
Despite these efforts, the nature of DLAs is still unclear \citep[for
  a review see][]{wolfegawiser05} and the interpretation of the large
observed velocity width of low-ionization species is not
unambiguous. In fact, the disk-like model of \citet{prochaskawolfe97}
in which DLAs are thick rotating disks with speed typical of a
present-day spiral galaxies and the alternative model based on the
assumption that DLAs are protogalactic clumps \citep{haehnelt98} seem
to be both viable. More precisely, \citet{haehnelt98} showed that
while the velocity width profiles of DLAS can be reproduced by
rotation in disks as well as protogalactic clumps, the velocity width
distribution cannot be reproduced by rotation in disks in the context
of a cold dark matter model.

In order to make significant progress in this field observational and
theoretical/numerical efforts are needed to tackle the problem under
many different aspects. Recently, the SDSS (Sloan Digital Sky Survey)
has allowed to measure to an unprecedented precision the statistical
properties of DLAs (incidence rates, column density distribution
function, etc.) over a wide redshift range
\citep{prochaska05}. Moreover, attempts to find the galaxy
counterparts of DLAs  represent the most promising way to understand
their nature \citep[e.g.][]{fynbo99,christensen07}.  High-resolution
spectroscopic studies have also played a crucial role since
low-ionization metal lines, that are supposed to be good tracers of the
neutral gas in DLAs, can be identified  shedding light on the DLAs
chemical and physical properties
\citep{matteucci97,calura03,vlaperoux05,vladilo06,prochaskawolfe08,vla08}. Here, we choose to focus
on the role of feedback in the form of galactic outflows and its
impact on the DLAs properties.

Recently, \citet{pontzen08} analysed several different hydrodynamical
simulations with and without feedback using an approximate
a-posteriori radiative transfer scheme. They matched most of the
observed DLAs properties apart from a tension with the observed
velocity widths. In their framework, haloes of virial masses between
$10^9$ and $10^{11}$ $h^{-1}M_{\rm \odot}$ were the main contributors
to the DLA cross-section. \citet{barnes08} with a semi-analytical model
claimed instead that in order to reproduce the velocity widths
distribution the cross-section of haloes less massive than $10^{10}$
$h^{-1}M_{\rm \odot}$ should be exponentially suppressed. Observations
of the HI distribution in the local universe made by \citet{zwaan08}
show that the link with galactic superwinds might be stronger than
expected and favoured this explanation instead of the disk-like model
to explain the observed velocity widths.  Also some correlation
properties of neutral hydrogen rich galaxies and HI absorbers in the
local universe seem to be better fit when galactic winds are taken
into account \citep{pbv08}. However, the situation is quite confusing
and potential problems for this interpretation are extensively
discussed in \citet{prochaska08}.

In this work, we extend the analysis performed recently by
\citet{nagamine07} that addressed the properties of DLAs in a
$\Lambda$CDM universe. The main differences can be summarized as
follows and will be more extensively discussed later: $i)$ we rely on
simulations that have on average a factor 10 better mass resolution;
$ii)$ we use a different version of the hydrodynamical code \gad that
incorporates the dependence of the radiative cooling function on the
global metallicity of the gas \citep[following][]{SD93} and a
self consistent metal enrichment model; $iii)$ we explore the effect
of varying the stellar initial mass function and most of the
parameters describing the wind model. As for the wind model we focus
mainly on energy-driven scenarios
but we allow for an extra simulation that is based on a
momentum-driven wind.  At least for the low-density IGM, this model
seems to better fit some observational properties \citep[
  e.g.][]{oppe06,oppe07}.  The goal is to see which DLAs properties
can be reproduced by the hydrodynamical simulations and to investigate
closely the impact of galactic winds both on the neutral hydrogen and
on the metal distribution around galactic environments in the high
redshift universe (mainly in the range $z=2-4$).

This work is intended as a preliminar quantitative attempt to match
some observed properties of the metal and neutral hydrogen
distribution using a self consistent chemo-dynamical code that has
been already tested for the intra-cluster medium \citep{T07}. After
having addressed some global properties and evolution of IGM
low-ionization species, we will focus on DLAs because their
statistical properties are well measured over a wide redshift range
and they could provide a useful benchmark for the physics implemented
in our simulations.

This paper is organized as follows. In Section 2 we describe our set
of simulations along with the two different galactic feedback
implementations used: energy--driven and momentum--driven winds. In
Section 3 we compare some global properties of the simulations: in
particular temperature and metallicity relations (Section 3.1), star
formation rates and evolution of ion species (Section 3.2). In Section
4 we focus on the neutral hydrogen distribution around galaxy-sized
haloes and we study DLAs properties like the cross-section (Section
4.3), the incidence rate (Section 4.4) and the column density
distribution function (Section 4.5). Section 5 is dedicated to the
simulated QSO spectra extracted from our simulations, to study the
velocity width distribution of low-ionization species (Section 5.1)
and the correlation between metallicity and velocity widths (Section
5.2). Finally in Section 6 we summarize our main results and we draw
some conclusions.

\section{The simulations}

We run a number of hydrodynamical simulations in order to explore the
effect of changing box-size, numerical resolution, stellar initial
mass function, prescription for energy feedback and nature of dark
matter.  We use a modified version of \gad, a parallel Tree-PM
Smoothed Particle Hydrodynamics (SPH) code \citep{springel2005}.  The
main modification consists in an accurate modelling of the chemical
evolution which allows us to follow the metal release from Type II and
Type Ia supernovae (SNII, SNIa), along with low and intermediate mass
stars (LIMS).  We refer to \citet{T07,fabjan07,borgani08} for more
details while we briefly outline in the following the most important
features of the chemical enrichment model.

The simulations cover a cosmological volume (with periodic boundary
conditions) filled with an equal number of dark matter and gas
particles.  The cosmological model chosen is a flat $\Lambda$CDM with
the following parameters: $\Omega_{\rm 0m}=0.24$, $\Omega_{\rm
  0b}=0.0413$, $\Omega_{\rm \Lambda}=0.76$, $n_{\rm s}=0.96$, $H_{\rm
  0}=73$ km/s/Mpc and $\sigma_{\rm 8}=0.8$, which are in agreement
with the latest results from large scale structure observables such as
the cosmic microwave background, weak lensing, the \lya forest and the
evolution of mass function of galaxy clusters
\citep{komatsu08,lesg08,vikhetal08}.  The input linear dark matter
power spectrum for the initial conditions has been generated at $z=99$
with {\small CMBFAST} \citep{seljakzalda96} and includes baryonic
acoustic oscillations. In one case only we change the initial linear
dark matter power spectrum by running a warm dark matter (WDM)
simulation with a modification (suppression at the small scales) of
the initial power spectrum to account for a thermal dark matter
particle of mass of 1.2 keV, which is in rough agreement with recent
results obtained from \lya high-resolution QSO spectra at redshifts
similar to those investigated here \citep{viel08}.

Radiative cooling and heating processes are included following the
implementation of \citet{KWH}. We assume a mean Ultra Violet
Background (UVB) produced by quasars and galaxies as given by
\citet{HM96}, with the heating rates multiplied by a factor $3.3$ in
order to better fit observational constraints on the temperature
evolution of the Intergalactic Medium (IGM) at high redshift. This
background gives naturally a $\Gamma_{\rm -12} \sim 0.8-1$ at low redshift
in agreement with recent observational measurements
\citep{bolton05,fg08}. Multiplying the heating rates by the factor
above (chosen empirically) results in a larger IGM temperature at the mean
density which cannot be reached by the standard hydrodynamical code
but aims at mimicking, at least in a phenomenological way, the
non-equilibrium ionization effects around reionization \citep[see for
  example][]{ricotti00,schaye00,Bol07}.  We follow self consistently
the evolution of the following elements: H, He, C, O, Mg, S, Si and
Fe.  The contribution of metals is included in the cooling function
adopting the tables of \citet{SD93}, that assume the solar value for
the relative abundances.  In this paper we use the solar metallicity
and element abundances given by \citet{asplund05}. 

The standard multiphase star formation criterion is used in which an
effective prescription for the inter-stellar medium (ISM) is
implemented \citep{springel2003}. In this effective model, a gas
particle is flagged as star forming whenever its density exceeds a
given density threshold, above which that particle is treated as
multiphase. With this prescription baryons are in the form either a
hot or a cold phase or in stars, thereby this density threshold marks
the onset of cold clouds formation. Following \citet{T07} we set the
threshold value to $\rho_{\rm th}=0.1$ cm$^{-3}$ in terms of the
number density of hydrogen atoms.

The neutral hydrogen fraction $f_{\rm HI}$ is associated to each gas
particle and is stored in each simulation snapshot. However, we follow
\cite{nagamine04} to assign a-posteriori a new mass in neutral
hydrogen to gas particles above the density threshold which reads:
\begin{eqnarray}
m_{\rm HI}\,\,\,\,\, =&f_{\rm HI}\, m_{\rm H}\;\;\; & (\rho <
\rho_{\rm th})\\ m_{\rm HI}\,\,\,\,\, =&f_{\rm c}\, m_{\rm H}\;\;\;
&(\rho \geq \rho_{\rm th})\;,
\end{eqnarray}
with $f_{\rm HI}$ the neutral hydrogen fraction that depends on the
UVB used, $m_{\rm H}$ the hydrogen mass of the particle ($f_{\rm HI}$
and $m_{\rm H}$ are determined self consistently inside the code),
$f_{\rm c}$ the fraction of mass in cold clouds and $\rho_{\rm th}$
the star formation threshold. Here $f_{\rm c}=\rho_{\rm c}/\rho$,
where $\rho_{\rm c}$ is the density of cold clouds and $\rho$ the
total (hot + cold) gas density. Individual molecular clouds cannot be
resolved at the resolution reachable in cosmological simulations, thus
$\rho_{\rm c}$ represents an average value computed over small regions
of the ISM.  We refer to \cite{springel2003} to better understand how
this cold fraction is related to the physics of the ISM.

Besides including different contributions from SNII, SNIa and LIMS,
our model of chemical evolution accounts for the age of various
stellar populations. Metals are thereby released over different
time-scales by stars of different mass. For the stellar yields we use:
SNIa -- \citet{thielemann03}; SNII -- \citet{woosleyweaver95}; LIMS --
\citet{vandenhoek97}. The mass-range for the SNII is $m > 8 M_{\rm
  \odot}$, while for the SNIa is $m < 8 M_{\rm \odot}$ with a binary
fraction of 10\%. We also adopt the lifetime function given by
\citet{padovanimatteucci93}.  Finally we use three distinct stellar
inital mass functions (IMFs): a Salpeter, a Kroupa and an
Arimoto-Yoshii IMF. Our reference choice is the functional form
proposed by \citet{salpeter55}: $\varphi(m) \propto m^{\rm -y}$, where
$\varphi(m)$ is the IMF by mass and $y=1.35$. Arimoto-Yoshii
\citep{ay87} IMF has $y_{\rm AY}=0.95$, while Kroupa IMF
\citep{Kroupa01} adopts a multi-slope approximation: $y_{\rm KR}=0.3$
for stellar mass $m < 0.5M_{\rm \odot}$, $y_{\rm KR}=1.2$ for
$0.5M_{\rm \odot} \leq m < 1M_{\rm \odot}$ and $y_{\rm KR}=1.7$ for $m
\geq 1M_{\rm \odot}$.

\begin{table*}
\label{tab:params}
\begin{tabular}{llccccccc}
\hline & Run & Size ($h^{-1}$Mpc) & \textit{N$_{\rm GAS}$}&
\textit{m$_{\rm GAS}$} ($h^{-1}$\msun) & soft. ($h^{-1}$kpc) & Wind
(km/s) & IMF & \textit{z$_{\rm f}$} \\ \hline & SW & 10 & $320^3$ &
$3.5 \times 10^5$ & 1.5 & 600 & Salpeter & 2.25 \\ & WW & 10 & $320^3$
& $3.5 \times 10^5$ & 1.5 & 100 & Salpeter & 2.25 \\ & MDW & 10 &
$320^3$ & $3.5 \times 10^5$ & 1.5 & $\sigma$-dependent$^{(a)}$ &
Salpeter & 2.25 \\ & SW$_{\rm KR}$ & 10 & $320^3$ & $3.5 \times 10^5$
& 1.5 & 600 & Kroupa & 2.25 \\ & SW$_{\rm AY}$ & 10 & $320^3$ & $3.5
\times 10^5$ & 1.5 & 600 & Arimoto-Yoshii & 2.25 \\ & SW$_{\rm
  10,448}$ & 10 & $448^3$ & $1.2 \times 10^5$ & 1 & 600 & Salpeter &
3.00 \\ & SW$_{\rm 10,256}$ & 10 & $256^3$ & $6.8 \times 10^5$ & 2 &
600 & Salpeter & 2.25 \\ & SW$_{\rm 20,512}$ & 20 & $512^3$ & $6.8
\times 10^5$ & 2 & 600 & Salpeter & 3.00 \\ & SW$_{\rm 5,320}$ & 5 &
$320^3$ & $4.3 \times 10^4$ & 0.75 & 600 & Salpeter & 3.00 \\ &
SW$_{\rm WDM}$$^{(b)}$ & 7.5 & $320^3$ & $1.5 \times 10^5$ & 0.75 &
600 & Salpeter & 3.00\\ \hline
\end{tabular}
\caption{Summary of the different runs. Column 1, run name: SW, Strong
  Winds, WW, Weak Winds and MDW, Momentum Driven Winds; column 2,
  comoving box
  size; column 3, number of gas particles;
  column 4, mass of gas particle; column 5, Plummer-equivalent
  comoving gravitational 
  softening; column 6, wind speed; column 7, Initial Mass
  Function (IMF) chosen (see text); column 8, redshift at which the simulation
  was stopped. $(a)$: $\sigma$ is the velocity dispersion of the halo
  that host the ``wind'' particle (see Section
  \ref{MDW_section}). $(b)$: WDM (Warm Dark Matter) simulation with a
  modification (suppression at the small scales) of the initial linear
  dark matter power spectrum to account for a thermal dark matter
  particle of mass 1.2 keV.}
\end{table*}

\subsection{Energy--driven winds}
\label{winds}

The implementation of galactic winds used is extensively described in
\citet{springel2003} however, since this paper addresses the
effects on the chemical enrichment of  winds blowing from high
density regions, we summarize here the major details of the original
model and the modifications that we made.

In the multiphase model of \citet{springel2003} gas contained in dark
matter haloes cool and settle into rotationally supported discs where
the baryons are converted into stars. However, in the star-forming
multiphase medium, it is plausible that not all of the hot gas in
supernova remnants remains confined to the disc: supernova
bubbles close to the surface of a star-forming region may break out of
a disc and a galactic-scale wind associated with star formation may
develop.

Basically, the wind mass-loss rate $\dot{M}_{\rm w}$ is assumed to be
proportional to the star formation rate $\dot{M}_{\rm \star}$
according to $\dot{M}_{\rm w}= \eta \dot{M}_{\rm \star}$, where $\eta$
is the wind efficiency.  Star-forming gas particles are then
stochastically selected according to their star formation rate to
become part of a blowing wind. Whenever a particle is uploaded to the
wind, it is decoupled from the hydrodynamics for a given period of
time (calculated as mentioned below) or till the density around drops
below a given fraction of the density threshold set for the onset of
the star formation, in order to effectively reach less dense
regions. In our case, this limiting density for decoupling is
$0.5\rho_{\rm th}$. This allows the wind particle to travel `freely'
up to few kpc until it has left the dense star-forming phase, without
directly affecting it: only outside the disc the hydrodynamical
interactions within the halo could stop the wind. Unlike in
\citet{springel2003}, we decide here to fix the velocity of the winds,
$v_{\rm w}$, instead of the fraction of the energy made available by
SNII explosions to power galactic ejecta. Thus, four parameters fully
specify the wind model: the wind efficiency $\eta$, the wind speed
$v_{\rm w}$, the wind free travel length $l_{\rm w}$ and the wind free
travel density factor $\delta_{\rm w}$.

The maximum allowed time for a wind particle to stay hydrodynamically
decoupled is then $t_{\rm dec} = l_{\rm w} / v_{\rm w}$.  The
parameter $l_{\rm w}$ has been introduced in order to prevent a gas
particle from getting trapped into the potential well of the
virialized halo and in order to effectively escape from the ISM, reach
the low density IGM and pollute it with metals. \citet{nagamine07}
showed that global DLAs properties are relatively insensitive to the
value of $l_{\rm w}$.

In this work, we will consider two values for the wind velocity:
$v_{\rm w}=100$ km/s (weak winds, WW) and $v_{\rm w}=600$ km/s (strong
winds, SW). In our implementation the parameter $\eta$ is kept fixed
to the value 2.  For a more quantitative description of the numerical
implementations made and for an application to the intra-cluster
medium physics we refer to the paper by \citet{T07}.

\subsection{Momentum--driven Winds}
\label{MDW_section}

In this Section we describe our implementation of the momentum--driven
galactic winds, which have been first implemented in simulations by
\citet{oppe06,oppe07}. These authors suggested that with this scheme
of winds simulations reproduce the statistics of CIV absorption in the
high--redshift IGM better than using the above energy-driven winds.

Observations of outflows from starbust galaxies have improved
considerably in recent years. \citet{Martin2005} found that the
terminal wind velocity scales roughly linearly with circular velocity,
with top winds speeds around three times the galaxy's circular
velocity. \citet{Rupkeetal2005} studied a large sample of luminous
infrared galaxies and found that, at least when combined with smaller
systems from \citet{Martin2005}, those trends continue to quite large
systems.

A feasible physical scenario for the wind driving mechanism is derived
by noting that the observed scaling are well explained by a
momentum--driven wind model such as that outlined by
\citet{Murrayetal2005}.  In such a scenario, the radiation pressure of
the starbust drives an outflow, possibly by transferring momentum to
an absorptive component (such as dust) that is hydrodynamically
coupled with the gas component, which is then dragged out of the
galaxy. Following \cite{oppe06, oppe07} we test a single model (very
close to their ``\textit{mzw}'' run) based on momentum--driven
winds. In such a model the wind speed scales as the galaxy velocity
dispersion $\sigma$, as observed by \citet{Martin2005}.  Since in
momentum--driven winds the amount of input momentum per unit star
formation is constant, this implies that the mass loading factor must
be inversely proportional to the velocity dispersion. We
therefore use the following relations:
\begin{equation}
\label{eqwind}
v_{\rm wind}=3\sigma\sqrt{f_{\rm L}-1};\ \ \ \ \eta=\frac{\sigma_{\rm 0}}{\sigma},
\end{equation}
where $f_{\rm L}$ is the luminosity factor in units of the galactic
Eddington luminosity (i.e. the critical luminosity necessary to expel
gas from the galaxy), $\sigma_{\rm 0}$ is the normalization of the
mass loading factor and we add an extra $2\sigma$ kick to get out of
the potential of the galaxy in order to simulate continuous pumping of
gas until it is well into the galactic halo. The functional form of
the luminosity factor includes a metallicity dependence for $f_{\rm L}$
owing to more UV photons output by lower--metallicity stellar
populations: $f_{\rm L}=f_{\rm L,\odot}\times10^{-0.0029(\log
  Z+9)^{2.5}+0.417694}$, where \citet{Martin2005} suggests: $f_{\rm
  L,\odot}\approx2$.  The mass loading factor controls star formation
at early times, so $\sigma_{\rm 0}$ can also be set by requiring a
match to the observed global star formation rate. Following
\citet{oppe07} we set $\sigma_{\rm 0}=150$ km/s. We determine $\sigma$
directly from the simulation only for the most massive haloes, while
for haloes with masses below $1.7\times 10^{9}$ $h^{-1}M_{\rm \odot}$
(i.e. formed by less than 10$^3$ dark matter particles) we use the
relation calibrated by \citet{evrardetal08}, using a variety of N-body
simulations:
\begin{equation}
  \sigma_{\rm DM}(M,z)=\sigma_{\rm DM,15}\left[\frac{h(z)M_{\rm 200}}{10^{15}M_{\rm \odot}}\right]^\alpha
  \ \ \ \ \textrm{km/s,}
\end{equation}
where $\sigma_{\rm DM,15} = 1082.9 \pm 4.0$ km/s is the normalization
at mass $10^{15} h^{-1} M_{\rm \odot}$, $\alpha=0.3361 \pm 0.0026$ is
the logarithmic slope and $M_{\rm 200}$ is defined as the total mass
within a sphere with mean interior density $3M_{\rm 200} / 4 \pi
r_{\rm 200}^3 = 200 \rho_{\rm c}(z)$ ($\rho_{\rm c}(z)$ is the
critical density at redshift $z$).  Even for this wind implementation
the particles are stochastically selected in the same way as for the
energy--driven scenario.\\

In Table I we summarize the main parameters of the cosmological
simulations performed including the mass associated to the gas
particles and the gravitational softening. All the simulations start
at redshift $z=99$. The reference simulations are SW (``Strong
Winds''), WW (``Weak Winds'') and MDW (``Momentum--Driven Winds'')
with a total of $2\times 320^3$ dark matter and gas particles in a 10
comoving $h^{-1}$Mpc box and Salpeter IMF.  Furthermore, SW$_{\rm KR}$
(Kroupa IMF) and SW$_{\rm AY}$ (Arimoto-Yoshii IMF) explore the
effects of a different IMF compared to the Salpeter one. Some
simulations are intended to explore resolution and box-size effects
like SW$_{\rm 10,448}$, SW$_{\rm 10,256}$, SW$_{\rm 20,512}$ and
SW$_{\rm 5,320}$. The most CPU-time consuming run is the SW$_{\rm
  20,512}$, which significantly increase the statistics of dark matter
haloes able to host DLAs. Finally, the SW$_{\rm WDM}$ aims at
exploring the effect of a modification in the initial linear dark
matter power spectrum.

\section{Global properties of the simulations}

\begin{figure*}
\includegraphics[width=5.8cm]{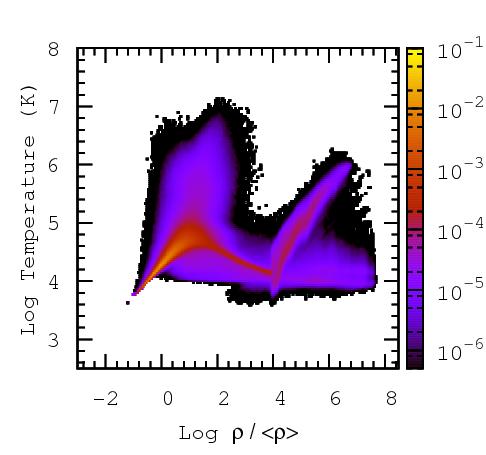}
\includegraphics[width=5.8cm]{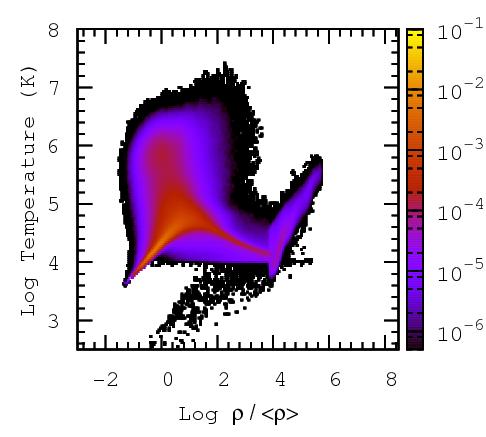}
\includegraphics[width=5.8cm]{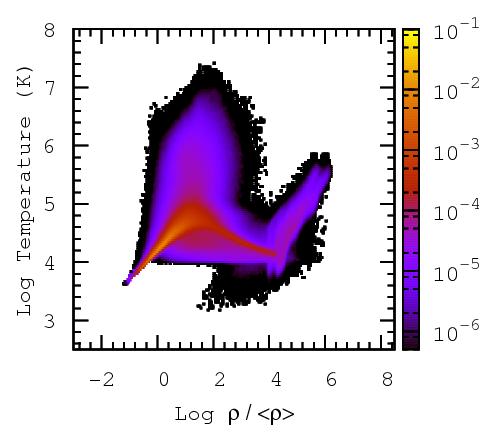}
\caption{Gas (IGM) temperature-density relation (without
  explicitly splitting the star-forming particles in the hot and the
  cold phases), for the WW (Weak galactic energy-driven Winds of 100
  km/s; left panel), SW (Strong galactic energy-driven Winds of 600
  km/s; middle panel) and MDW (Momentum-Driven galactic Winds; right
  panel) simulations at $z=3$. The vertical bar indicates the gas mass
  fraction.}
\label{fig:rt_sw}
\end{figure*}

\begin{figure*}
\includegraphics[width=5.8cm]{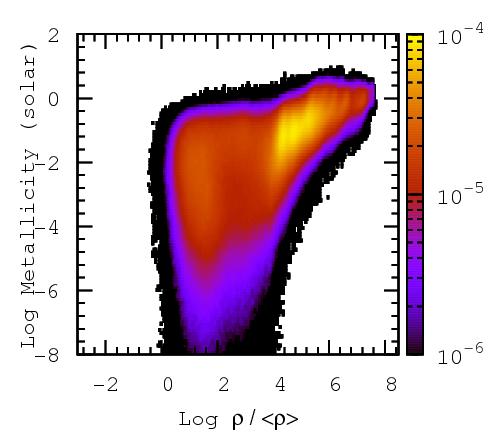}
\includegraphics[width=5.8cm]{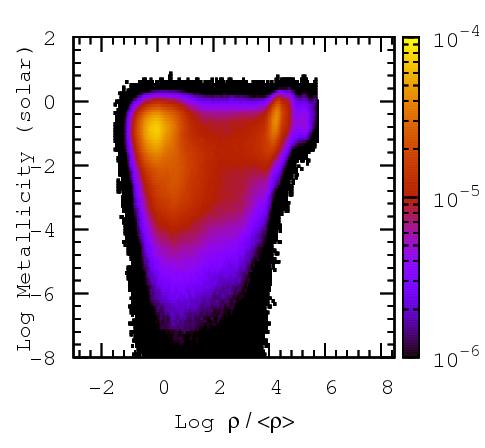}
\includegraphics[width=5.8cm]{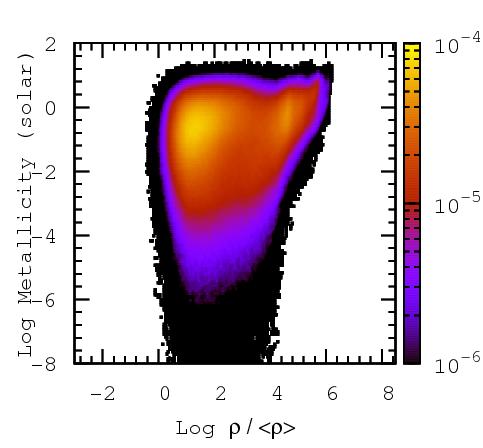}
\caption{Gas (IGM) metallicity-density relation (without explicitly
  splitting the star-forming particles in the hot and the cold
  phases), for the WW (Weak galactic Winds of 100 km/s; left panel),
  SW (Strong galactic Winds of 600 km/s; middle panel) and MDW
  (Momentum-Driven galactic Winds; right panel) simulations at $z=3$.
  The vertical bar indicates the gas mass fraction.}
\label{fig:rz_sw}
\end{figure*}

\subsection{IGM temperature and metallicity relations at $z=3$}

In this Section we investigate two global properties of the IGM, namely
its thermal and chemical state, focussing on the differences between
the SW, WW and MDW runs at $z=3$. A more detailed analysis of
global IGM properties down to small redshifts is beyond the scope of
this paper and will be presented in a future work.

In Figures \ref{fig:rt_sw} and \ref{fig:rz_sw} we show the IGM
temperature-density relation and the metallicity-density relation for
the WW (left panel), SW (middle panel) and MDW (right panel)
simulations. In these figures it is shown an ``effective'' temperature
for the star-forming particles determined as the mean temperature
weighted by the contribution in mass of the cold and the hot phases
(see Section 2). In all cases the phase diagrams are color coded
according to the gas mass fraction, as indicated by the vertical
bars. For all the three runs a significant fraction in mass (about
10\%) of the IGM resides in the tight power-law relation $T=T_{\rm 0}
(1+\delta)^{\gamma-1}$ at around the mean density. Gas at this density
is usually responsible for \lya forest absorption. At $z=3$ the IGM
has a temperature at the mean density of about $T_{\rm 0}=15000$ K and
the slope $\gamma$ of the temperature-density relation is around
1.6. In our simulations we assume photoionization equilibrium and
under this assumption it is difficult to get values of $\gamma<1.6$,
which however are in better agreement with the recent measurements of
the \lya probability distribution function \citep{bolton08}. An IGM
fraction between $10^{-4}$ and $10^{-3}$ is in the form of
shock-heated gas at temperatures higher than $10^5$ K. The tail of
cold gas at large densities resides in the inner parts of the haloes
while at even larger densities, $>10^4 \langle\rho\rangle$, and
temperature, $>10^4$ K, is associated to star-forming regions. We note
a plume of cold and overdense gas particles that are below $10^4$ K
and carry a fraction of mass of the order $10^{-6}$: these are gas
particles that are in the wind phase. After these particles enter the
wind regime they become hydrodynamically decoupled for a certain
period of time (see Section 2.1). During this period they travel
`freely' towards regions in which both the density and the temperature
are lower. As a consequence, when these particles re-couple to the
hydrodynamics, their temperature drop because of the adiabatic
expansion.

\begin{figure*}
 \includegraphics[width=8.5cm,height=8cm]{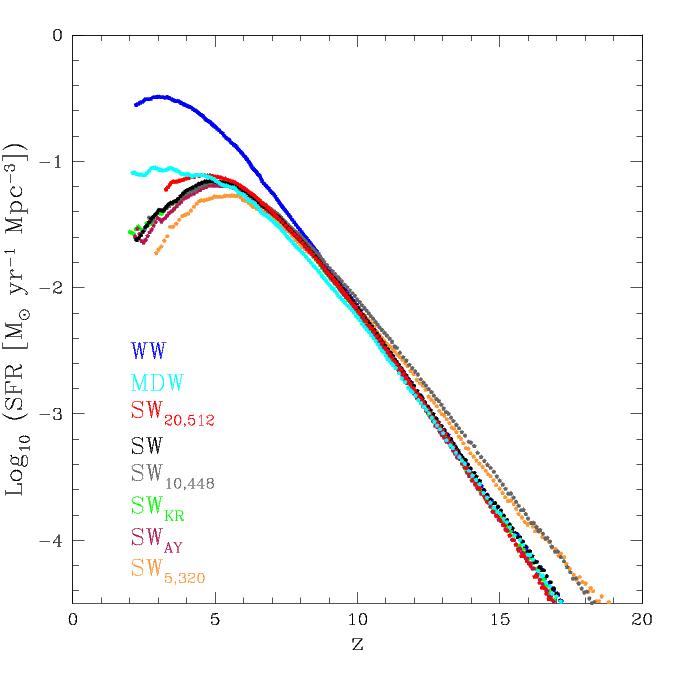}
 \includegraphics[width=8.5cm,height=8cm]{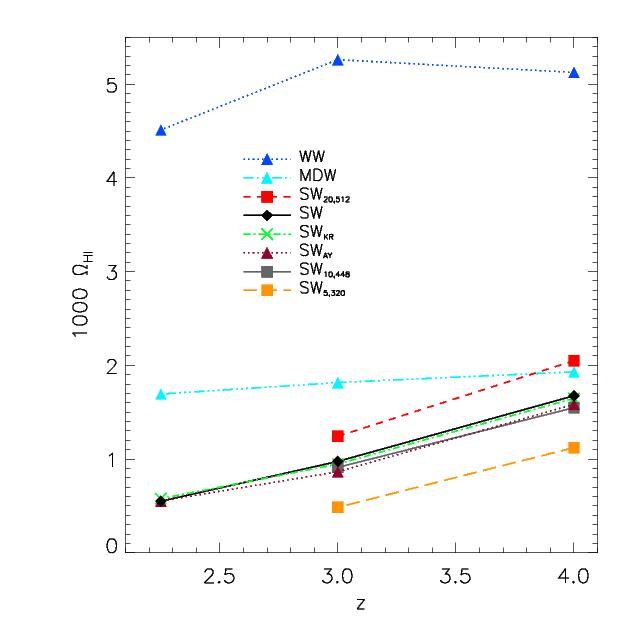}
 \caption{{\it {Left Panel}}: cosmic star formation rate (SFR) for
   some of the hydrodynamical simulations of Table I. {\it Right
     Panel}: evolution of the total $\Omega_{\rm HI}$ ($\times 1000$)
   as a function of redshift for some of the hydrodynamical
   simulations of Table I.}
 \label{fig:sfrOmHI}
 \end{figure*}

In Figure \ref{fig:rz_sw} we show the metallicity-density relation for
the three simulations. In the SW case the gas that has been enriched
spans a wide range of densities and can attain values that are very
close to solar. Also gas that is below the mean density appears to be
metal enriched at a high level in this simulation. The bulk ($10^{-4}$
in mass fraction) of the IGM is either in the form of gas at the mean
density or in the form of very dense gas at about
$10^4\langle\rho\rangle$ with metallicities of about 0.1 \zsun.

All the previous findings are particularly interesting when the WW
(left panel) and SW (middle panel) runs are compared. For the WW
temperature-density relation it is clear that the amount of
shock-heated gas is signifcantly reduced and the region of the diagram
at temperatures between $10^5$ and $10^7$ K is less populated. Thereby, strong
galactic winds with speed of order 600 km/s heat the IGM
significantly, while weaker winds with speed of order 100 km/s are
less effective in doing this. The gas particles in the wind phase at
$T<10^4K$ are now fewer than in the SW case and at higher
overdensity. Moreover, the star-forming high density tail is
considerably more extended than in SW case confirming the less
efficient feedback of WW in suppressing star formation. In contrast
with the SW simulation, in the WW case there is slightly less
metal-enriched underdense gas, while the metallicity gradient with
density is steeper. Higher values of metallicities are now reached,
which attain super-solar values at very high density. In the WW case
the bulk of the enriched gas shifts from the mean density values of
the SW case to a region which is about $10^4$ times denser than the
mean.

We stress that the difference between the SW and WW simulation is only
in the wind speed which is 600 and 100 km/s, respectively, while all
the other parameters have been kept fixed.  These values of the wind
velocity have been chosen in order to embrace values suggested by
observational studies of HI and CIV-galaxy correlation and \lya forest
in close QSO pairs \citep{adelberger05,rauch05}, and observations of
interstellar lines in ultraluminous infrared galaxies
\citep{Martin2005}.

 The right panels of Figures \ref{fig:rt_sw} and \ref{fig:rz_sw} show
 results for the MDW simulation. Looking at the temperature-density
 diagram one can easily see a general trend that is intermediate
 between those of the SW and WW runs. For example, the region
 associated to wind particles (cold and overdense particles below
 $10^4$ K) is less pronounced with respect to SW but not as negligible
 as the WW run.  However the metallicity-density relation is quite
 different from the energy-driven implementation. In particular the
 metallicity reaches values slightly higher than SW and WW for the
 whole range of densities and most importantly there is much more
 enriched material around the mean density. This result suggests that
 MDW is more efficient in polluting the IGM above and around the mean
 density: this is due to the fact that, unlike for SW and WW,
 in MDW even small haloes contribute significantly to the
 enrichment. In fact, such small haloes have small velocity dispersion
 and correspondingly greater loading factors (see Eq.  (\ref{eqwind}))
 for their winds.

\subsection{Star formation rates and evolution of ion species}
\label{sec:sfrprop}
In this Section we analyse the star formation rates (SFR) for the
different simulations and the evolution of the neutral hydrogen
content and of two ion species, CIV and OVI, that are usually observed
in absorption in QSO spectra even at higher redshifts than those
explored here \citep[e.g.][]{becker08,ryan-weber09}.

In the left panel of Figure \ref{fig:sfrOmHI} we show the total star
formation rate of the simulated volume as a function of redshift.  The
WW simulation has nearly a factor 10 higher star formation rate
compared to the others at $z\sim 3$. This means that the feedback
mechanism induced by strong galactic winds is effective in paupering
the metal-rich star-forming gas significantly decreasing the star
formation rate. The SW$_{\rm 20,512}$ and SW$_{\rm 5,320}$ embrace all
the SW runs and this is due to cosmic variance effects. We note
negligible differences between the Salpeter IMF and the other two IMFs
used (Kroupa and Arimoto-Yoshii), at least at the relatively high
redshifts considered here. We underline, however, that we neglect the
effect of assuming different IMFs on the observationally inferred
cosmic star formation rate; this means that we do not change the star
formation efficiency as it would be required in order to match the
observables when the number of massive stars per unit mass of formed
stars changes. That is because here we are more interested in the
chemical and energetic effect of the IMF. We also point out the
intermediate trend of the MDW, in between SW and WW. At very high
redshift ($z>10$) the SW$_{\rm 5,320}$ and SW$_{\rm 10,448}$ show a
higher star formation with respect to the other runs. This is due to
the improved resolution of these simulations that can resolve higher
densities at earlier times. The star formation rate of the SW$_{\rm
  10,448}$ at lower redshift agrees very well (in fact the black and
the grey curves are nearly indistinguishable) with that of the SW
simulation, which has the same box size but less particles. This
confirms the numerical convergence of our simulations.

\begin{figure*}
\centering
\includegraphics[width=14.5cm, height=7cm]{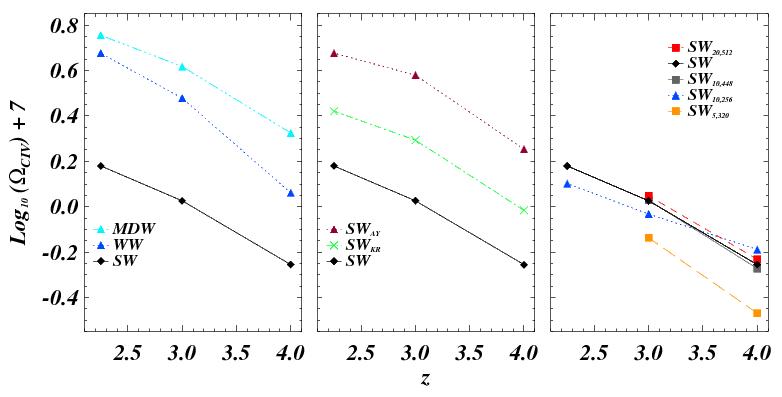}
\includegraphics[width=14.5cm, height=7cm]{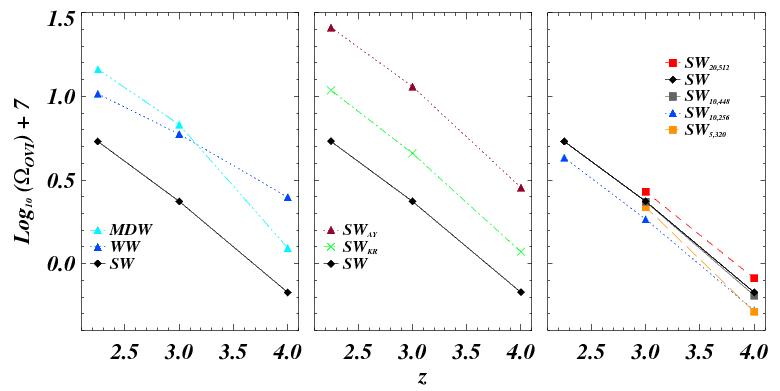}

\caption{Evolution of the total $\Omega_{\rm CIV}$ (upper panels) and
  the total $\Omega_{\rm OVI}$ (lower panels) as a function of
  redshift for some of the hydrodynamical simulations of Table I. {\it
    {Left Panels}}: effect of different wind strengths and
  implementations. {\it {Middle Panels}}: effect of different
  IMFs. {\it {Right Panels}}: resolution tests.}
\label{fig:CIVOVI}
\end{figure*}

The behaviour shown in the star formation rate is very similar in
nature to that of the total neutral hydrogen evolution of Figure
\ref{fig:sfrOmHI} (right panel). Here we plot $\Omega_{\rm HI}$, which
is defined as the contribution of neutral hydrogen to the total
critical density (the values have been multiplied by 1000). The
neutral hydrogen fraction is followed in the simulation self
consistently with the assumed average UV background and not rescaled
a-posteriori using a different UV background.  Note again that the
neutral hydrogen content of the WW simulation is about a factor 5
higher than that of the SW and this is due to the fact that in the WW
simulation the gas is colder and more concentrated in the potential
wells of galaxies and thereby is significantly more neutral. Even in
this case the SW$_{\rm 20,512}$ and SW$_{\rm 5,320}$ simulations
embrace all the SW runs as for the star formation rate plot. The
SW$_{\rm KR}$ and SW$_{\rm AY}$ agree well with the SW confirming that
feedback and resolution/box-size effects have more impact on the
simulations than effects due to the particular choice of the IMF as
long as the stellar spectra are not self consistently taken into
account.  The difference between WW and SW simulations is somewhat
more pronounced than that found by \citet{nagamine04} and this is
probably due to the fact that our simulations include metal cooling,
thereby increasing the neutral hydrogen amount in the haloes (we will
come back to this point later). The MDW is in between SW and WW and
again the SW$_{\rm 10,448}$ is in good agreement with the SW,
demonstrating that, at least for these quantities, numerical
convergence has been reached.

At the end of Section \ref{sec:coldensdistr} we compute the DLAs
contribution to the total neutral hydrogen content ($\Omega_{\rm
  DLA}$) and we compare it with the results of \citet{pontzen08} and
the observational estimates of \citet{prochaska05} and
\citet{peroux05}.

As a further check, in Figure \ref{fig:CIVOVI} we plot the evolution
of the contribution to the total density of two of the ions that trace
the high redshift IGM at relatively low density and for which there
are some observational constraints \citep{schaye03,aguirre07}: CIV
($\lambda\lambda$ 1548.204, 1550.781 \AA) in the upper panels and OVI
($\lambda\lambda$ 1031.927, 1037.616 \AA) in the lower
panels. Although in the rest of the paper we will focus on the
distribution of metals around DLA systems, we would like to address
briefly the evolution of ionization species in the IGM as a whole. We
use the redshift outputs at $z=2.25,3,4$ and we extract the values for
$\Omega_{\rm CIV}$ and $\Omega_{\rm OVI}$ summing over all the gas
particles. In order to obtain the ionization fraction for the two
elements it is necessary to multiply the abundance of a given metal,
carried by each particle, by its ionization fraction that depends on
density and temperature.  We use the {\small CLOUDY} code
\citep{ferland} to compute a-posteriori the relevant fractions for
each gas particles. We choose the {\small HM05} option in {\small
  CLOUDY}, which consists of a UVB made by QSOs and galaxies with a
10\% photon escape fraction and which is in agreement with other
observational constraints \citep{bolton05}.

In Figure \ref{fig:CIVOVI} we explicitly show how different effects
impact on $\Omega_{\rm CIV}$ and $\Omega_{\rm OVI}$ : in the left
panels we compare simulation with different wind strength and
implementation (SW, WW and MDW); in the middle panels we test the
effect of different IMF (SW, SW$_{\rm KR}$ and SW$_{\rm AY}$); in the
right panels we present resolution tests for all the SW runs (SW,
SW$_{\rm 5,320}$, SW$_{\rm 10,256}$, SW$_{\rm 10,448}$ and SW$_{\rm
  20,512}$). Here we discuss each panel of the figure:
\begin{itemize}
\item {\it {Left Panels (feedback physics)}}: for the SW simulations
  the total amount of the two ions increase by a factor of about 3
  between $z=4$ and $z=2.25$.  The WW simulation contain $\sim 2.5$
  times more OVI and CIV compared to the SW case. The reason is that
  with such weak winds the metals remain very close to the dense
  environments around galaxies and cannot reach the low-density IGM in
  contrast with the SW case \citep[see for example][]{Theuns02}. In
  these regions close to galaxies the ionization fractions are
  tipically larger than in the voids. The MDW CIV and OVI evolution is
  similar to the energy driven implementations although it has a
  higher normalization than SW and WW: this is due to the fact that in
  the former implementation the winds are more efficient in enriching
  the IGM around the mean density with metals and this produces larger
  values of $\Omega_{\rm CIV}$ and $\Omega_{\rm OVI}$ reached at
  $z=2.25$.
\item {\it {Middle Panels (IMF)}}: The trend in the redshift
  dependence of $\Omega_{\rm CIV}$ and $\Omega_{\rm OVI}$ is the same
  for SW, SW$_{\rm KR}$ and SW$_{\rm AY}$, while there are differences
  in the normalization. The Kroupa and Arimoto-Yoshii IMFs result in
  values for the CIV and OVI density that are respectively $\sim$1.5
  and $\sim$3 times higher than the standard Salpeter case. This is
  due to the fact that Salpeter IMF results in an excess of low mass
  stars, Kroupa IMF produces a smaller number of massive stars than
  the other two but twice as many stars in the range of mass
  $0.3M_{\rm \odot}<m<3M_{\rm \odot}$, while with the Arimoto-Yoshii
  (or ``top-heavy'') IMF there is a larger contribution from massive
  stars \citep[see Figure 3 in][]{T07}. For these reasons, the
  SW$_{\rm AY}$ run produces more oxygen and carbon than the other
  simulations, while the SW run (with Salpeter IMF) is less efficient
  in producing these metal species and SW$_{\rm KR}$ is in between the
  other two.  Interestingly the MDW run with Salpeter IMF
  (triple-dot-dashed cyan lines in the left panels) produces more CIV
  than the SW$_{\rm AY}$, while the amount of OVI is comparable,
  confirming the high efficiency of momentum driven winds in enriching
  the IGM.
\item {\it {Right Panels (resolution tests)}}: the aim of these two
  panels is to show that our analysis is robust against box-size and
  resolution effects. We plot $\Omega_{\rm CIV}$ and $\Omega_{\rm
    OVI}$ for all the SW runs and it is clearly evident the
  convergence of the results (especially for the $\Omega_{\rm OVI}$
  and the SW$_{\rm 10,448}$ simulation). As we already found for the
  SFR and the $\Omega_{\rm HI}$, the SW$_{\rm 20,512}$ and SW$_{\rm
    5,320}$ simulations embrace all the SW runs.
\end{itemize}

A comparison of these findings with those by \cite{oppe06,oppe07}
shows that the evolution of $\Omega_{\rm CIV}$ is somewhat faster and
the normalization higher than what they find. In particular our MDW is
even more discrepant than SW and WW with respect to the analogous
momentum-driven winds simulation (``\textit{mzw}'') by \cite{oppe06}.
However, it is difficult to compare properly given the different
resolutions, box-sizes, feedback implementations and details on
chemical evolution model used. \\

In the last part of this Section we briefly discuss the other
simulations described in Table I. The simulations SW, SW$_{\rm
  10,448}$ and SW$_{\rm 10,256}$ are characterised by the same box
size but different number of particles, so that they allow us to carry
out a test of stability of our results against numerical
resolution. As already mentioned, for the SW$_{\rm 10,448}$ the star
formation rate trend at high redshift is similar to that of SW$_{\rm
  5,320}$ (the simulation with the highest resolution), while it
agrees at lower redshift with SW, which has the same box size. Instead
the SW$_{\rm 10,256}$ has lower SFR (not plotted) than SW at high
redshift because, for a given box-size, a lower number of particles
means a lower number of structure resolved in the simulation and
thereby a lower SFR. As for the evolution of $\Omega_{\rm HI}$,
$\Omega_{\rm CIV}$ and $\Omega_{\rm OVI}$, the SW$_{\rm 10,448}$
follows the results of the SW almost perfectly. This is also true for
other analysis made in this paper (phase diagrams, haloes properties,
etc.), thus further confirming the numerical convergence of our
results on these simulations properties. The run SW$_{\rm WDM}$ has a
different initial power spectrum $P(k)$ respect all others
simulations: power is suppressed at small scales due to free streaming
of the dark matter particles that in this case are ``warm'' and have
mass equal to 1.2 keV (corresponding to a suppression scale of around
50 comoving $h^{-1}$kpc). As a consequence, the star formation rate
density at high redshift is significantly reduced with respect to the
CDM case, with a correspondingly smaller number of haloes having mass
comparable or smaller than the free-streaming mass scale. We run this
simulation in order to test whether the modified $P(k)$ could produce
haloes that better fit the column density distribution function for
small values of $N_{\rm HI}$, that will be discussed in Section
\ref{sec:coldensdistr}.

\section{The neutral hydrogen distribution}

In this Section we focus on the properties of the neutral hydrogen
around the simulated galactic haloes. The aim is to study DLAs
properties and statistics extracted from our simulations and to
compare them with the latest observational data and the simulation
results from other groups \citep{nagamine04,nagamine07,pontzen08}.

\subsection{Identifying haloes}
\begin{figure}
\includegraphics[width=8cm]{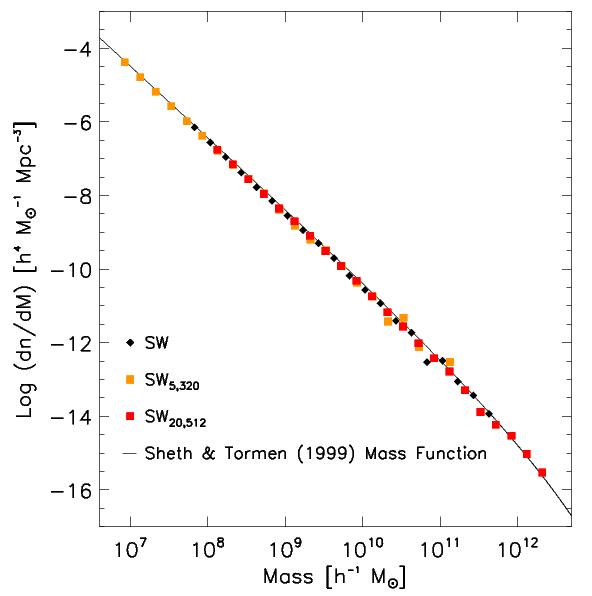}
\caption{Differential halo mass function (number of haloes of a given
  mass per unit mass [$h^{-1}M_{\rm \odot}$] and per unit volume [Mpc$^3$/$h^3$]) for runs SW, SW$_{\rm 5,320}$, SW$_{\rm 20,512}$ at
  redshift $z=3$, compared with \citet{shethtormen99} prediction.}
\label{fig:PS}
\end{figure}

We identify haloes in the simulations by running a parallel
Friends-of-Friends (FoF) algorithm with a linking length which is 0.2
times the dark matter mean interparticle spacing. The number of haloes
of a given mass per unit mass and per unit volume as a function of
their mass is shown in Figure \ref{fig:PS} for the SW, SW$_{\rm
  5,320}$ and SW$_{\rm 20,512}$ runs at redshift $z=3$, along with the
\citet{shethtormen99} mass function, so as to test at the same time
the effect of resolution and of box size. Again, the SW$_{\rm 20,512}$
and SW$_{\rm 5,320}$ embrace the reference case SW: the mass function
of SW$_{\rm 5,320}$ extends to small halo masses due to its better
resolution, while SW$_{\rm 20,512}$ produce more massive haloes due to
its larger box size. In our largest simulation we have three haloes
above $10^{12} h^{-1}M_{\rm \odot}$ and about 1000 haloes of masses
above $10^{10} h^{-1}M_{\rm \odot}$ at $z=3$.

\begin{figure}
\includegraphics[width=7.5cm,height=7.5cm]{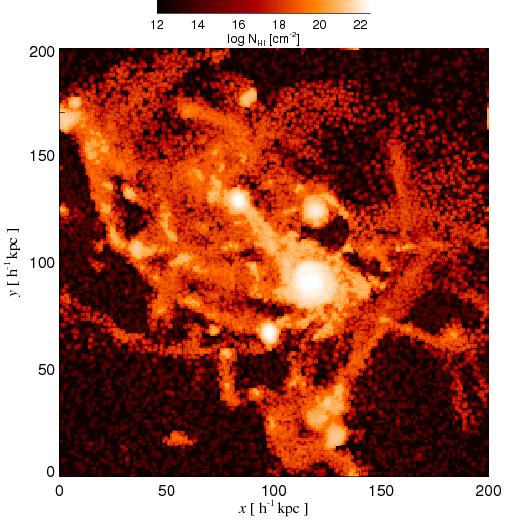}
\includegraphics[width=7.5cm,height=7.5cm]{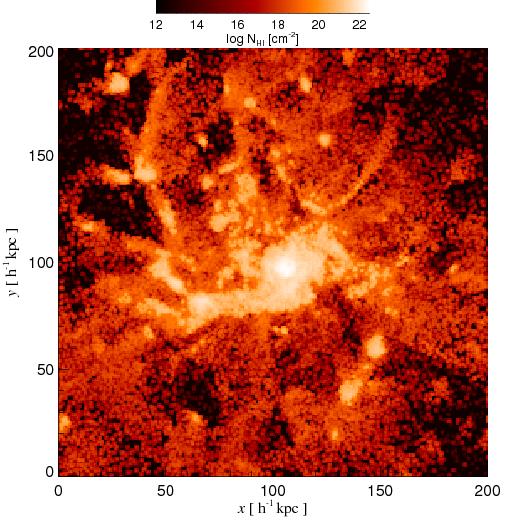}
\includegraphics[width=7.5cm,height=7.5cm]{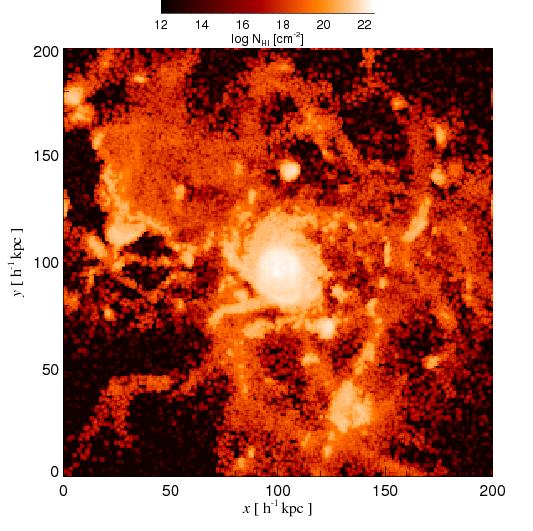}
\caption{HI column density maps in a slice around the same massive
  halo in the WW (upper panel), SW (middle panel) and MDW (lower
  panel) runs at $z=3$.}
\label{fig:dla_256}
\end{figure}

We follow the analysis made by \citet{nagamine04,nagamine07} to
realize a mock DLA sample: after having identified the haloes and
their center of mass, we interpolate with a TSC (Triangular Shaped
Cloud) algorithm the comoving neutral hydrogen mass density around the
center of mass of each halo on a cubic grid; then we `collapse' the
grid along a random direction and we obtain a set of neutral hydrogen
column densities for each halo. Thus the column density reads:
\begin{equation}
N_{\rm HI}=\Sigma_{\rm i}\,\rho_{\rm i,HI} \, \epsilon/m_{\rm
  p}(1+z)^2\;,
\end{equation}
with $m_{\rm p}$ the proton mass and $\epsilon=l/n_{\rm grid}$ the
linear dimension of the single grid cell. Here $l$ is the size of the
box around the halo and $n_{\rm grid}$ is the number of grid points.

Tipically, for the most massive haloes, we use cubes of size 200
comoving $h^{-1}$kpc with $32^3$ grid points
($\epsilon=6.25h^{-1}$kpc). In such a way, we increase the DLA total
redshift path and we sample $32^2 \times N_{\rm haloes}$ HI column
densities along lines-of-sight per simulated box.

\begin{figure*}
\includegraphics[width=18cm,height=9cm]{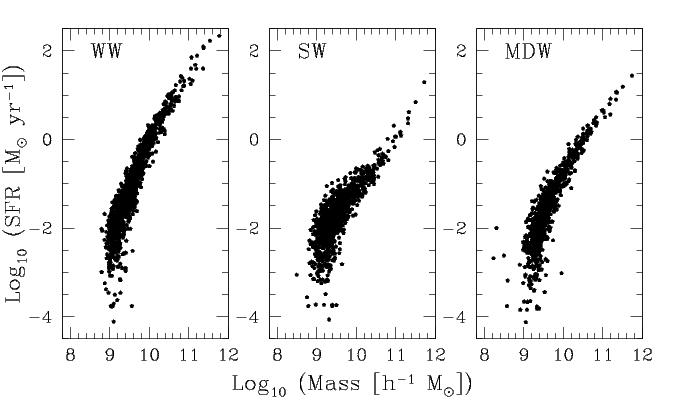}
\caption{Star formation rates (in $M_{\rm \odot}$ per year) plotted as
  a function of halo mass for SW, WW and MDW runs at $z=3$.}
\label{fig:sfr_h}
\end{figure*}

\begin{figure*}
\includegraphics[width=18cm,height=9cm]{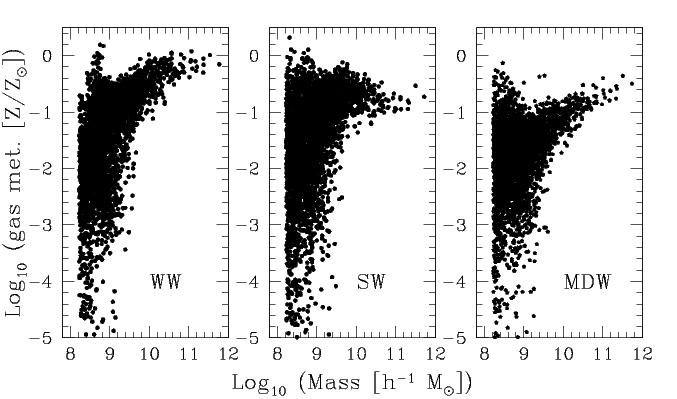}
\caption{Mean total metallicity in solar unit plotted as a function
  of halo mass for SW, WW and MDW runs at $z=3$.}
\label{fig:met_h}
\end{figure*}

\begin{figure*}
\includegraphics[width=18cm,height=9cm]{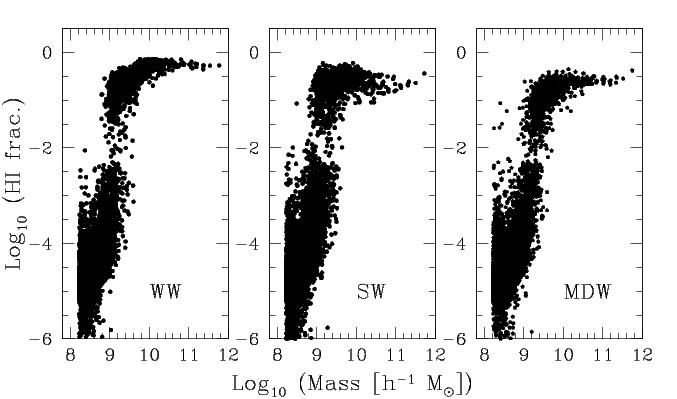}
\caption{Mean neutral fraction of hydrogen (HI/H) plotted as a
  function of halo mass for SW, WW and MDW runs at $z=3$.}
\label{fig:hi_h}
\end{figure*}

We have carried out some tests changing the number of grid points in
order to study the effect of the sampling size on the neutral hydrogen
distribution. The choice of the grid points number is crucial because
too many points produce a sampling size below the resolution of the
simulation and consequently an ``oversampling'' of the HI mass density
with large statistical fluctuations, while too few points produce a
smooth statistic which could not be representative of the real density
field. At the end we found that the best compromise was to use $32^3$
grid points, corresponding to $\epsilon$ $\magcir 4.5 \times
softening$ $length$ (which is also the typical value of the SPH
smoothing length in the outskirts of the haloes), differently from
\citet{nagamine04}, who choose instead $\epsilon \approx softening$
$length$.

Figure \ref{fig:dla_256} shows the HI column density maps extracted as
explained above, but with a finer grid subdivision of 256 points, for
the same massive halo in the WW (upper panel), SW (middle panel) and
MDW (lower panel) runs at $z=3$. The HI density at each pixel has been
projected along the line-of-sight in the z direction. In the figure it
is visible the effect of the winds: in the WW run (upper panel) high
column density gas is more concentrated inside the central halo and
inside some substructures. Also the column density values reached are
higher than for the other runs. In the SW run (middle panel), the gas
is more spread around the central haloes and the
substructures. Finally if we consider HI column densities above the
DLA limit of $N_{\rm HI} = 2\times 10^{20}$ cm$^{-2}$ we see that the
central halo in the MDW run has the largest cross-section. We discuss
further about this point in Section \ref{sec:cross}.

\subsection{Properties of the haloes}

For the haloes identified with the FoF algorithm we compute mean
quantities that could be relevant for the following analysis, such as
the star formation rate inside each halo, the mass-weighted mean total
metallicity and the mass-weighted mean neutral fraction of hydrogen
(HI/H).  We plot in Figures \ref{fig:sfr_h}, \ref{fig:met_h} and
\ref{fig:hi_h} our findings at redshift $z=3$, only for haloes having
mass greater than $2\times 10^8h^{-1}$ \msun, that are resolved with
at least 100 dark matter particles. The properties of the haloes below
$10^9h^{-1}$ \msun, resolved with $\sim$ 1000 particles, should not be
trusted at a quantitative level, but we prefer to show them in order
to appreciate the increase in the scatter at low masses.

For the most massive haloes, the WW simulation shows very high star
formation rates that are even a factor 100 larger than that of the
corresponding haloes in the SW run: this is expected due to the
smaller efficiency of the weak winds in suppressing star formation
(see the left panel of Figure \ref{fig:sfrOmHI}). The MDW run reaches
nearly the same SFR values of the SW for the most massive haloes,
while for intermediate mass haloes ($10^{9.5-11}h^{-1}M_{\rm \odot}$)
the MDW haloes have higher SFR than SW ones. For the smallest haloes
of around $10^9h^{-1} M_{\rm \odot}$ the SW run shows a slightly
larger scatter in the SFR values. Overall, the bulk of the haloes, of
masses between $10^9$ and $10^{10}h^{-1}$ \msun, that are likely to
host DLA systems, have star formation rates of about 0.1
\msun/year. For haloes above $10^{10}h^{-1}M_{\rm \odot}$ the star
formation rates are different between the models and usually around
1-30 \msun/year.  These star formation rates are in general agreement
with those of the population of faint \lya emitters found recently by
\citet{rauch08}, in which a link between the \lya emitters and the
DLAs is suggested.

For the same reasons metallicities of haloes in the WW run are higher
than in the SW and MDW simulations (see Figure
\ref{fig:met_h}). Comparing this result with Figure \ref{fig:CIVOVI},
one can derive the following general picture: in the WW run metals
remain locked inside the haloes (or stay close to the haloes), while
in SW and MDW they are able to reach the IGM and enrich it: this is
particularly true for the less massive haloes in the MDW. In fact, the
MDW small mass haloes have in general metallicity lower than the
corresponding SW and WW haloes. The wind implementation is thereby
very effective in devoiding the galaxies of star forming cold gas that
is enriched at a level of about 0.1 \zsun.  The most important result
of this panel is however that the different wind implementations show
different mass-metallicity relations especially for haloes of masses
above $10^{9.5}h^{-1} M_{\rm \odot}$: while WW and MDW show a
correlation (although with different amplitude), the SW results seem
to produce very little correlation between mass and metallicity or
possibly a weak anti-correlation.  This different metallicity pattern
in haloes could be important when compared with observation
\citep[e.g.][]{maiolino08} and could possibly allow to discriminate
between different feedback scenarios.

Finally we note that the largest metallicities appear to occur in
haloes of small masses. There are two different motivations for this
to happen. The first motivation is numerical: as we mentioned at the
beginning of this Section, both these small haloes and the wind model
are not well resolved, thereby their large metallicities are, at least
partially, the result of some numerical artifact. Furthermore, we also
checked that there is a physical motivation: these small haloes are
strongly affected both by self enrichment at early epochs and by the
metal enriched winds blowing from bigger haloes close to
them. Addressing quantitatively these issues would require more
numerical work, which is beyond the scope of this work.

In Figure \ref{fig:hi_h} we show the mass--weighted mean neutral
fraction of hydrogen (HI/H) inside haloes, a quantity which is closely
related to the DLA properties. The general trend reflects those of the
right panel of Figure \ref{fig:sfrOmHI}: for a given halo mass the WW
simulation displays the highest value, while the SW and the MDW have
comparable values. In all the panels, there are present two different
sets of values: above and below HI/H $\sim$0.01. The first is
associated to the most massive haloes and the second to the least
massive ones. This is due to the fact that the most massive haloes
contains many particles above the density threshold $\rho_{\rm th}$
for which the neutral hydrogen fraction is equal to the fraction of
mass in cold clouds $f_{\rm c}$ (see Eqs. (1) and (2) in Section 2),
boosting in this way the neutral hydrogen content of these
haloes. Less massive haloes have instead neutral fractions that are
set by the physical conditions of the gas and by the UVB background
and are typically more ionized ($f_{\rm HI} \ll f_{\rm c}$) than the
most massive haloes.

\begin{table}
\begin{center}
\label{tab:crossect}
\begin{tabular}{llcccccc}
\hline & Redshift & Run & slope $\alpha$ & $\beta$ &  \\ \hline \hline

 & & SW & 0.57 & 3.83 & \\ 
 & $z=4$ & WW & 0.46 & 3.79 & \\ 
 & & MDW & 0.69 & 4.11 & \\ \hline 

 & & SW & 0.77 & 3.75 & \\ 
 & $z=3$ & WW & 0.49 & 3.61 & \\ 
 & & MDW & 0.85 & 4.06 & \\ \hline 

 & & SW & 0.62 & 2.99 & \\ 
 &$z=2.25$ & WW & 0.52 & 3.45 & \\ 
 & & MDW & 0.92 & 3.88 & \\ \hline 
\end{tabular}
\end{center}
\caption{Fitting parameters $\alpha$ and $\beta$ of
  Eq. (\ref{eq:fitcross}) for runs SW, WW and MDW at redshit $z=4$, 3
  and 2.25.}
\end{table}

\subsection{The DLA cross-section}
\label{sec:cross}
\begin{figure*}
\includegraphics[width=16cm]{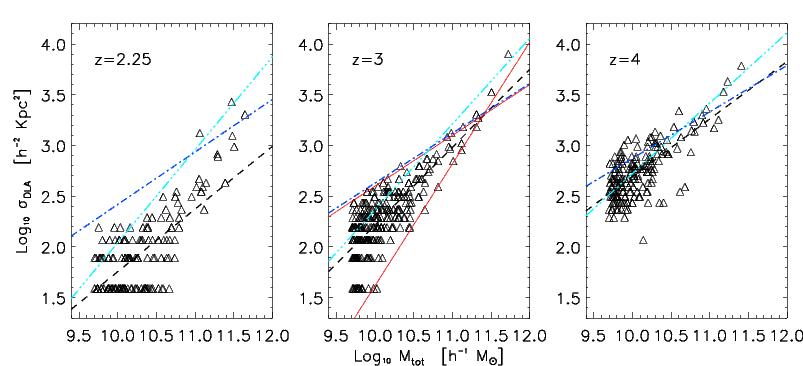}
\caption{Triangles represent the DLA cross-sections (in comoving
  units) as a function of the total halo mass for the SW run at
  $z=2.25,3,4$. The overplotted lines show power-law fits, of the form
  $\log \sigma_{\rm DLA}= \alpha (\log M_{\rm tot} - 12) + \beta$, to
  the data points for the SW itself (dashed black line), WW
  (dot-dashed blue line) and MDW (triple-dot-dashed cyan line)
  runs. The continuous red lines in the middle panel fit the upper and
  lower envelopes of the SW distribution at $z=3$.}
\label{fig:cross}
\end{figure*}

For each halo of total mass $M_{\rm tot}$ we derive the DLA
cross-sections, $\sigma_{\rm DLA}$ (in comoving units), by
selecting and summing up the area of all the cells with a column
density (determined as explained in Section 4.1) above $10^{20.3}$
cm$^{-2}$. Next we fit a power-law relation of the form:
\begin{equation}
\label{eq:fitcross}
\log \sigma_{\rm DLA}= \alpha (\log M_{\rm tot} - 12) + \beta \;,
\end{equation}
for all the simulations made. Results are shown in Figure
\ref{fig:cross} where we plot the cross-sections as extracted from the
SW at $z=2.25,3,4$ with overplotted fits for SW itself (dashed black
line), WW (blue dot-dashed line) and MDW (cyan triple-dot-dashed
line). Fitting parameters $\alpha$ and $\beta$ for each of these runs
are show in Table II. The continuous red lines in the middle panel fit
the upper and lower envelopes of the SW distribution at $z=3$ and were
drawn with the aim of highlighting the scatter in the distribution of
cross-sections expecially at low masses. The parameters of the upper
envelope line are $\alpha=0.5$ and $\beta=3.6$ (very similar to the WW
case), while those of the lower envelope line are $\alpha=1.2$ and
$\beta=4.02$. In the next section we discuss the effect of taking into
account this scatter in a very conservative way.

It is clear that for the WW simulation the cross-sections are larger
than in the SW case for the small mass haloes: this is due to the fact
that the wind is more effective in expelling cold gas from small mass
haloes than from the most massive ones. For the most massive haloes
the content of cold gas is much larger and this trend is inverted: the
WW run has a smaller cross-section than in the SW and MDW cases
because the cold gas is located very close to the dense galactic
environments. The MDW trend reflects the fact that momentum-driven
winds are more directly related to the halo properties: smaller mass
haloes have greater loading factor and winds become very efficient as
in SW case. On the contrary, for massive haloes the MD winds become
less efficient: the cold blob of gas around the halo has a more
regular shape than in the SW case and determines a larger
cross-section than the SW.  To sum up, as one can see from Figure 6,
the qualitative trends are the following: for WW the gas is
concentrated in the center of the haloes; the SW expels gas in a quite
violent way and the shells of expanding material fragment; the MDW
blows less gas but in a more homogeneous way than SW.

A comparison of the results from the SW run with those obtained by
\citet{nagamine04,nagamine07} at similar resolution shows that we
obtain on average a 20 per cent shallower slopes and 5\% smaller
$\beta$ parameters: this is likely to be due to the faster wind speed
adopted for the strong wind case (600 km/s vs. 484 km/s) that will
probably slightly reduce the normalization $\beta$ devoiding galaxies
of cold gas. Furthermore, the fact that our simulations have the metal
cooling implemented could result in a larger cross-section for smaller
haloes because the quantity of cold gas increases, while for the most
massive haloes the metal cooling enhances the amount of stars at the
expenses of cold gas. These two effects determine a shallower slope
for the cross-section fitting. In the WW case the run should be
compared with the P3 run of \citet{nagamine04,nagamine07} even though
the wind speed adopted in this study is 100 km/s vs. 242 km/s and the
mass resolution is about 10 times better, but even in this case the
trend is confirmed and we find a shallower slope, of about 40 per
cent, and a smaller normalization value (by $\sim 10$ per cent).

In Figure \ref{fig:cross} it is clearly visible, especially at redshift
$z=2.25$, a discretization in the cross-section values, more
pronounced than in \citet{nagamine04,nagamine07}. This is due to our
final choice of the linear dimension of the single grid cell
$\epsilon$ which is somewhat larger than that adopted by \citet{nagamine04}.

\subsection{The incidence rate of DLA systems}

Having obtained the mean relation for the DLA cross-section as a
function of halo mass from the previous Section, we are now able to
calculate the cumulative number of DLAs per unit redshift (or rate of
incidence) using the equation:
\begin{equation}
\label{eqdNdz}
\frac{dN_{\rm DLA}}{dz}(>M,z)=\frac{dr}{dz}\int_M^{\infty}n_{\rm h}(M',z)\,\sigma_{\rm DLA}(M',z)\,dM',
\end{equation}
where $n_{\rm h}(M,z)$ is the \citet{shethtormen99} dark matter halo
mass function and $dr/dz=c/H_{\rm 0}\sqrt{\Omega_{\rm
    m}(1+z)^3+\Omega_{\rm \Lambda}}$. Following \citet{nagamine04}, we
use this equation in order not to be sensitive to dark matter haloes
with masses below the resolution limit of the simulation. This is a
common problem when one tries to compute the number density of DLAs
based on a cosmological simulation that does not resolve all small
mass haloes that may host a DLA: the underlying assumption in the rest
of the paper is that haloes below $10^9h^{-1}M_{\rm \odot}$ are not
able to produce DLA systems. Moreover small box size simulations
cannot produce very massive haloes. To overcome these limitations in
Eq. (\ref{eqdNdz}) we convolved the \citet{shethtormen99} halo mass
function with the measured relationship between DLA cross-section and
halo mass, thereby correcting for incompleteness in the resolved halo
abundance of our simulations. In doing this we extrapolate the
power-law fit of Eq. (\ref{eq:fitcross}) both at high and low
masses. This relation presents an intrinsic scatter: looking at Figure
\ref{fig:cross} one can note that a given $\sigma_{\rm DLA}$ value
corresponds to different halo masses, especially at low
cross-sections. As mentioned in the previous section we check the
impact of this scatter fitting the upper and lower envelopes of the
cross-sections distribution of the SW run at $z=3$.

\begin{figure}
\includegraphics[width=8cm]{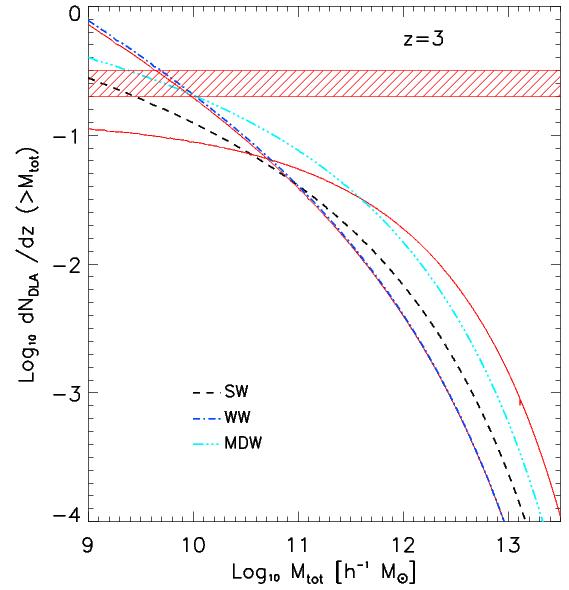}
\caption{Cumulative abundance of DLAs per unit redshift as a function
  of total halo mass for the SW (dashed black line), WW (dot-dashed
  blue line) and MDW (triple-dot-dashed cyan line) runs at redshift
  $z=3$. The continuous red lines show the results considering the
  scatter (in a conservative way) in the cross-section vs mass
  relation for the SW run.  The red shaded region indicates the
  observed cumulative DLA abundance of \citet{prochaska05} from SDSS
  data.}
\label{fig:dndz}
\end{figure}

In Figure \ref{fig:dndz} we show $dN_{\rm DLA}/dz$ for SW, WW and MDW
runs at $z=3$. The three different runs have identical initial power
spectrum and cosmological parameters so they all have the same
theoretical dark matter halo mass function. Therefore the differences
in Figure \ref{fig:dndz} reflects what we found in previuos Section
about DLAs cross-sections (see in particular the central panel in
Figure \ref{fig:cross}). In fact at low masses WW simulation produce
haloes with cross-sections $\sim$3 times higher than SW and MDW, and
correspondingly $dN_{\rm DLA}/dz$ reaches higher values. At greater
masses the trend is inverted and WW curve stays below the other
two. As we expected $dN_{\rm DLA}/dz$ curve for MDW simulation is
always above the SW one and well above WW one in the high mass
tail. The shaded region shows the observational estimate
log($dN/dz$)=$-0.60 \pm$0.10 at $z=3$ recovered by \citet{nagamine07}
from the observational results based on SDSS QSO spectra of
\citet{prochaska05}. Finally, the continuous red lines show the
results considering the scatter in the cross-section vs mass relation
for the SW run. In one case (where the curve is almost equal to that
of the WW case) one would need only haloes with mass greater than
$\sim10^{9.7}h^{-1}M_{\rm \odot}$ to fit the $dN_{\rm DLA}/dz$
statistic, while in the other case one should go down to much less
massive haloes, that are not well resolved by our simulations.

Our findings for the DLA abundance per unit redshift are slightly
different than those obtained by \citet{nagamine07}: moving to low
masses their distribution tend to flatten, while ours are somewhat
steeper and this is due both to the different values of the
cross-section, the different dimension of the grid cells used, and the
different cosmological parameters for the linear dark matter power
spectrum (amplitude and slope). As a result, at low masses, the
$dN_{\rm DLA}/dz$ values in our simulations are about a factor $\sim
2$ greater than the corresponding ones of \citet{nagamine07}, while at
high mass the trend is inverted and our $dN_{\rm DLA}/dz$ decreases
somewhat more rapidly than Nagamine et al. results. To summarize, our
results reproduce the observational data of \citet{prochaska05}
slightly better than those of \citet{nagamine07} which slightly
underpredict $dN_{\rm DLA}/dz$ in the range of masses considered.

As far as the redshift evolution of DLA rate of incidence is concerned,
the behaviour of the three different simulations reflects that of
Figure \ref{fig:cross}. For the sake of brevity we do not show all the
plots, but at $z=4$ the decreasing in $dN_{\rm DLA}/dz$ at
high masses is much more pronounced than at $z=3$, due to a lack in
massive haloes that have not already formed. At $z=2.25$ both WW and
MDW curves stay always above the SW one and are more flattened than before
down to low masses. The overall trend with redshift is the same of the
left panel in Figure 6 of \citet{nagamine04}.

\begin{figure*}
\includegraphics[width=8cm]{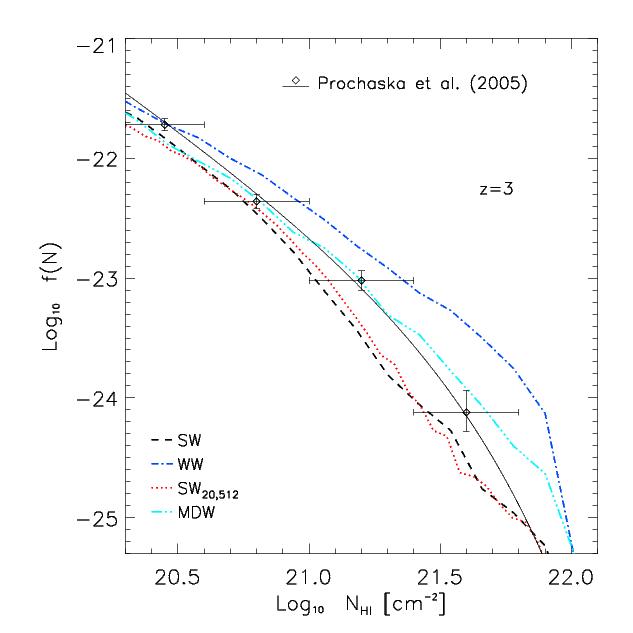}
\includegraphics[width=8cm]{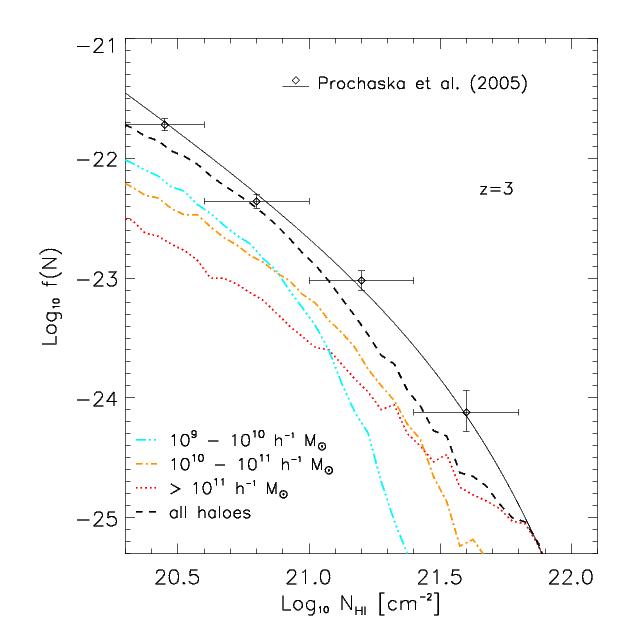}
\caption{{\it Left Panel}: HI column density distribution function at
  $z=3$ for the SW (dashed black line), WW (dot-dashed blue line),
  SW$_{\rm 20,512}$ (dotted red line) and MDW (triple-dot-dashed cyan
  line) runs.  {\it Right Panel}: Contribution to the HI column
  density distribution function from haloes of different mass in the
  SW$_{\rm 20,512}$ run at $z=3$. Cyan triple dot-dashed line refers
  to haloes with mass in the range 10$^9-10^{10}h^{-1}M_{\rm \odot}$,
  orange dot-dashed line refers to haloes with mass in the range
  10$^{10}-10^{11}h^{-1}M_{\rm \odot}$, red dotted line refers to
  haloes with mass greater than 10$^{11}h^{-1}M_{\rm \odot}$ while
  black dashed line refers to all haloes. In both the panels the
  overplotted black diamonds and the solid line show the data points
  and the fit of \citet{prochaska05}.}
\label{fig:fn}
\end{figure*}

\begin{figure*}
\includegraphics[width=8cm]{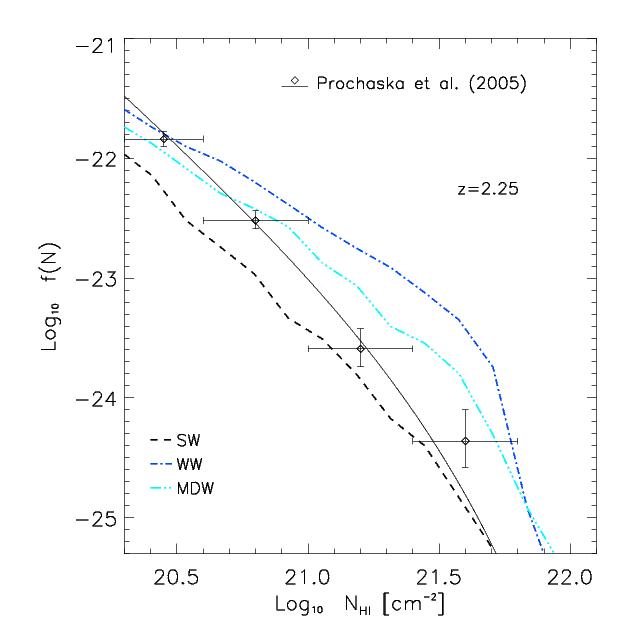}
\includegraphics[width=8cm]{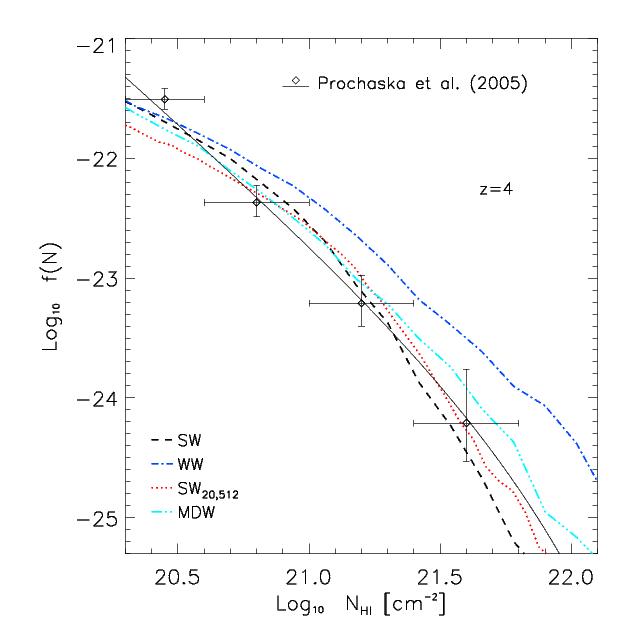}
\caption{{\it Left Panel}: HI column density distribution function at
  $z=2.25$ for the SW (dashed black line), WW (dot-dashed blue line)
  and MDW (triple-dot-dashed cyan line) runs. Overplotted black
  diamonds and solid line show the data points and the fit of
  \citet{prochaska05}. Run SW$_{\rm 20,512}$ is not present here
  because this simulation was stopped at $z=3$. {\it Right Panel}: the
  same as the left panel but at redshift $z=4$. Dotted red line refers
  to the SW$_{\rm 20,512}$ run.}
\label{fig:fn_z2.25}
\end{figure*}

\subsection{The column density distribution function}
\label{sec:coldensdistr}
In this Section we investigate the column density distribution
function for DLAs usually plotted in the form of $f(N)$, where
$f(N,X)dNdX$ is the number of DLAs with HI column density in the range
$[N,N+dN]$ and absorption distances in the interval $[X,X+dX]$. The
absorption distance is given by $X(z)=\int_0^z(1+z')^2H_0/H(z')\,dz'$.

In the left panel of Figure \ref{fig:fn} we show $f(N)$ for the SW
(dashed black line), WW (dot-dashed blue line), SW$_{\rm 20,512}$
(dotted red line) and MDW (triple-dot-dashed cyan line) runs at
$z=3$. The data from SDSS are overplotted along with the
$\Gamma$-function fit (black diamonds and solid line) of
\citet{prochaska05}. SW$_{\rm 20,512}$ is plotted to test the effect
of different resolution and box size and its trend is similar to
SW. Both are in quite good agreement with data especially at low and
high column densities. Instead WW overpredicts the distribution
function at large $N_{\rm HI}$ as was already found by
\citet{nagamine04}. Finally, MDW agrees extremely well with the data
down to the smallest $N_{\rm HI}$ values.  Comparing with the results
of \citet{nagamine04,nagamine07} we find a better agreement with the
observational data both for the SW and MDW, especially at low column
density and at $z=3$.

In the right panel Figure \ref{fig:fn} we split the contribution of
the different haloes to the total $f(N)$ for the SW$_{\rm 20,512}$ at
$z=3$. We use the SW$_{\rm 20,512}$ run because the largest box size
improves the statistics on the haloes. Looking at the figure one can
easily see that haloes with masses above $10^{11}h^{-1}M_{\rm \odot}$
(dotted red line) contribute primarily to the large column densities,
while the smallest haloes with masses in the range
$10^{9}$-$10^{10}h^{-1}M_{\rm \odot}$ (triple-dot-dashed cyan line)
contribute significantly to the lower column densities, below $N_{\rm
  HI} \approx 10^{20.8}$ cm$^{-2}$. This is not surprising since on
average more massive haloes have higher gas densities and
corrispondingly produce larger HI column densities values.

In Figure \ref{fig:fn_z2.25} we show the same as Figure \ref{fig:fn}
but at redshift $z=2.25$ (left panel) and $z=4$ (right panel), to
check the redshift evolution of the column density distribution
function.  In the left panel, the results for SW$_{\rm 20,512}$ are
not shown because this simulation ended at $z=3$. SW fits well the
observational data for HI column densities larger than $10^{21}$
cm$^{-2}$, but at lower column densities there is a discrepancy of
about a factor 3. WW and MDW overproduce $f(N)$ at high $N_{\rm HI}$,
but there is a better agreement with data at log$N_{\rm HI} < 20.7$,
especially for the MDW run.  At redshift $z=4$ (right panel) SW,
SW$_{\rm 20,512}$ and also MDW match very well the data while WW does
not. The fact that MDW fits the data as well as SW at $z=4$ is due to
the fact that high redshift momentum-driven winds behave quite
similarly to energy-driven ones, while at lower redshift the two
models tend to differ significantly in terms of velocities and loading
factors. Thus, we conclude that the different wind implementations of
the galactic wind feedback show distinct predictions for the redshift
evolution, and in general the differences become larger when moving to
$z\sim 2$. At the end the feedback model that reproduce better the
data is the MDW model.

At the end of Section \ref{sec:sfrprop} we briefly discussed the
SW$_{\rm WDM}$ simulation. We run this simulation motivated by the
poorer fit to observational data for the systems of column densities
between $10^{20.3-20.8}$ cm$^{-2}$ that was found in
\citet{nagamine04,nagamine07} and \citet{pontzen08}.  Our reference
runs (especially the SW) show a similar flattening trend and
underproduce the number of these systems, by a smaller amount when
compared to \cite{nagamine07}.  The idea is that using a different
linear power sprectrum we could possibly modify the halo mass function
and this could impact on the column density distribution function as
well. We perform the same analysis of the other runs for SW$_{\rm
  WDM}$, but we find no statistical significant difference for the
whole column density range. The smaller number of dark matter haloes,
compared to $\Lambda$CDM cosmology, is thus compensated by an increase
in the cross-section, since in WDM the haloes are usually less
concentrated.  Thereby we decided not to consider this run anymore and
we conclude that it appears unlikely that this statistic could be
better fit by invoking modifications of the linear dark matter power
spectrum such as warm dark matter.

Finally, we compute the total neutral gas mass in DLAs using:
\begin{equation}
\label{om_dla}
  \Omega_{\rm DLA}(z)=\frac{m_{\rm p}\,H_{\rm 0}}{c\,f_{\rm HI}\,\rho_{\rm {c},0}}\int_{10^{20.3}}^{N_{\rm max}}f(N_{\rm HI},X)N_{\rm HI}\,dN_{\rm HI},
\end{equation}
where $m_{\rm p}$ is the proton mass, $f_{\rm HI}$ is the neutral
hydrogen fraction of the gas and $\rho_{\rm c,0}$ is the critical
density at redshift $z=0$. The integration limit goes from $10^{20.3}$
cm$^{-2}$ (the lower column density limit for a system to be
identified as DLA) to $N_{\rm max}=10^{21.75}$ cm$^{-2}$, this latter
chosen to compare with the results of \citet{pontzen08}.

\begin{figure}
\includegraphics[width=8cm]{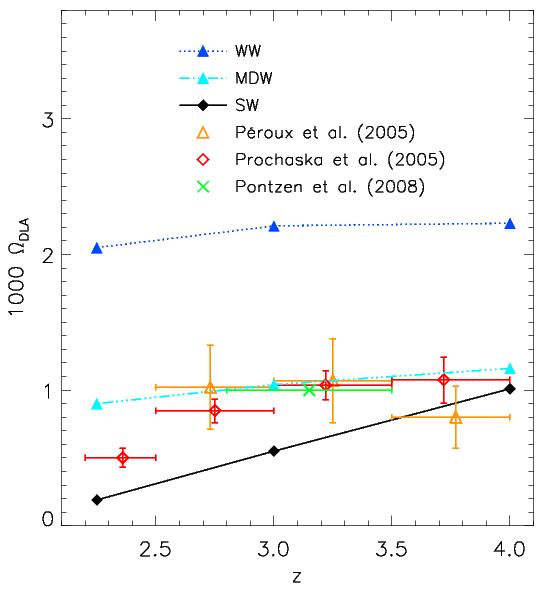}
\caption{Redshift evolution of $\Omega_{\rm DLA}$ for the SW (black
  solid line), WW (blue dotted line) and MDW (cyan triple-dot-dashed
  line) runs. Overplotted are the result of \citet{pontzen08} (green
  cross) and the observational estimate of \citet{prochaska05} (red
  diamonds) and \citet{peroux05} (orange triangles).}
\label{fig:OmDLA}
\end{figure}

In Figure \ref{fig:OmDLA} we show the evolution with redshift of
$\Omega_{\rm DLA}$ for the SW (black solid line), WW (blue dotted
line) and MDW (cyan triple-dot-dashed line) runs, along with the
result of \citet{pontzen08} (green cross) and the observational
estimate of \citet{prochaska05} (red diamonds) and \citet{peroux05}
(orange triangles). The different simulations' trends reflect those
plotted in the right panel of Figure \ref{fig:sfrOmHI}: the WW run produces
the largest $\Omega_{\rm DLA}$ value while SW the smallest. The amount of
neutral hydrogen in DLAs is about a factor two smaller than the total
neutral hydrogen in the simulated volume. The MDW is in good agreement
with the value found by \citet{pontzen08} ($1.0\times 10^{-3}$ in the
redshift range $2.8<z<3.5$) and with the data by \citet{peroux05},
while at redshift below $z\sim3$ it is slightly larger than the recent
measurement made by \citet{prochaska05} using DLAs in SDSS spectra.

\begin{figure*}
\includegraphics[width=17cm, height=18cm]{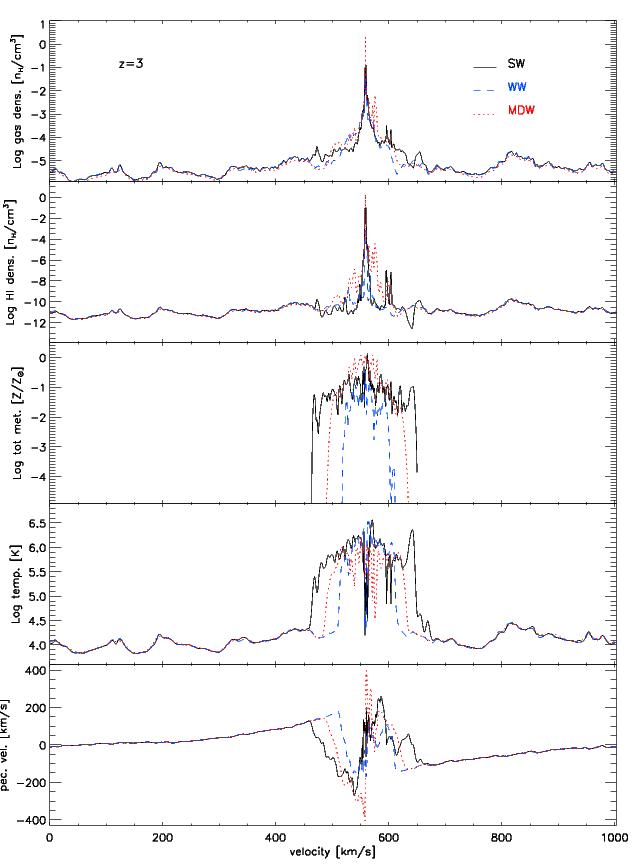}
\caption{Comparison of physical quantities extracted from three
  simulated LOSs passing through the center of mass of the second most
  massive halo for the SW (solid black line), WW (dashed blue line)
  and MDW (dotted red line) runs at $z=3$. From top to bottom the
  figure shows: total gas density [n$_{\rm H}/$cm$^{3}$], HI density
  [n$_{\rm H}/$cm$^{3}$], total metallicity (in solar unit),
  temperature [K] and peculiar velocity field [km/s].}
\label{fig:los}
\end{figure*}

So far we did not comment much about self-shielding effects. A full
treatment of the self-shielding would require radiative transfer and
this again would be an approximate (usually a-posteriori) scheme.  We
decided to rely on the multiphase ISM model to implictly account for
self-shielding motivated by the two following facts: $i)$ many
observational properties of DLAs have been reproduced in such a way
\citep{katz96,nagamine04,nagamine07}; $ii)$ the recent results by
\citet{pontzen08} seem to suggest that even when applying crude
radiative transfer approximations the main properties of DLAs do not
change significantly. However, to better check the impact of this
criterion, we also apply a further approximation that is useful in
order to decouple the star formation from the multiphase prescription
and assume that the gas particles above densities of $n_{\rm
  H}=10^{-2}$ cm$^{-3}$ are fully neutral
\citep[following][]{haehnelt98}. We recomputed the column density
distribution function for a few simulations and found that the
differences are not large. All the column density distribution
function increase by a fixed overall amount of about 0.2 dex and are
still in broad agreement with the observations. Since we believe that
our multiphase ISM model is more refined and more physically motivated
than this criterion we decided not to present results for this second
simpler assumption.

\section{Simulated QSO spectra}

For each simulation performed we have extracted several physical
quantities interpolated along lines-of-sight (LOSs) through the box,
following the procedure of \citet{theunsetal98} (Appendix A4). The
optical depths of the simulated QSO spectra are drawn in redshift
space taking into account the effect of the IGM peculiar velocities
along the line-of-sight, $v_{\rm pec,\parallel}$. As already mentioned
in Section 2, our simulations follow self consistently the evolution
of H, He, C, O, Mg, S, Si and Fe. Using the {\small CLOUDY} code
\citep{ferland}, we then determine the ionization fractions for some
ions that trace the high redshift IGM: CIV ($\lambda\lambda$ 1548.204,
1550.781 \AA), OVI ($\lambda\lambda$ 1031.927, 1037.616 \AA) and SiII
($\lambda$ 1526.707 \AA), with the possibility to extend the analysis
to many others. The simulated flux of a given ion transition at the
redshift-space coordinate $u$ (in km/s) is $F(u)=\exp[-\tau(u)]$ with:
\be \tau(u)={\sigma_{\rm 0,X} ~c\over H(z)} \int_{-\infty}^{\infty}
dx\, n_{\rm X}(x) ~{\cal V}\left[u-x-v_{\rm pec,\parallel}^{\rm
    IGM}(x),\,b(x)\right]dx \label{eq1} \;, \ee where $\sigma_{\rm
  0,X}$ is the cross-section of the particular ion transition, $H(z)$
is the Hubble constant at redshift $z$, $x$ is the real-space
coordinate (in km s$^{-1}$), $b=(2k_{\rm B}T/m_{\rm X}c^2)^{1/2}$ is
the velocity dispersion in units of $c$, ${\cal V}$ is the Voigt
profile.

\begin{figure*}
\includegraphics[width=8cm]{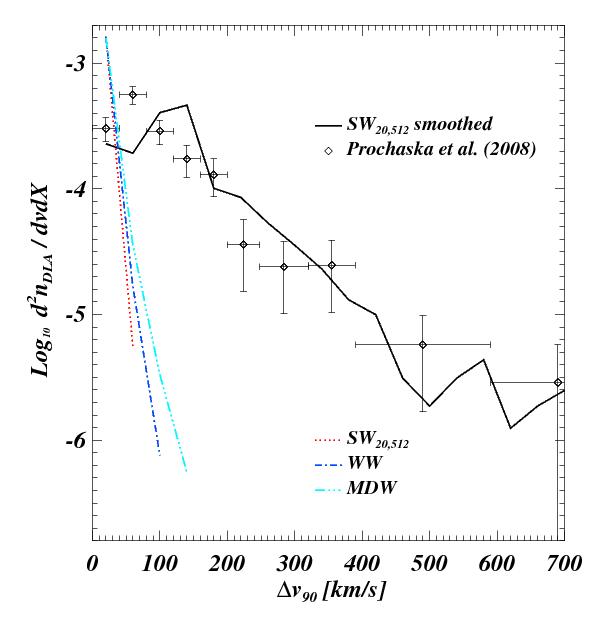}
\includegraphics[width=8cm]{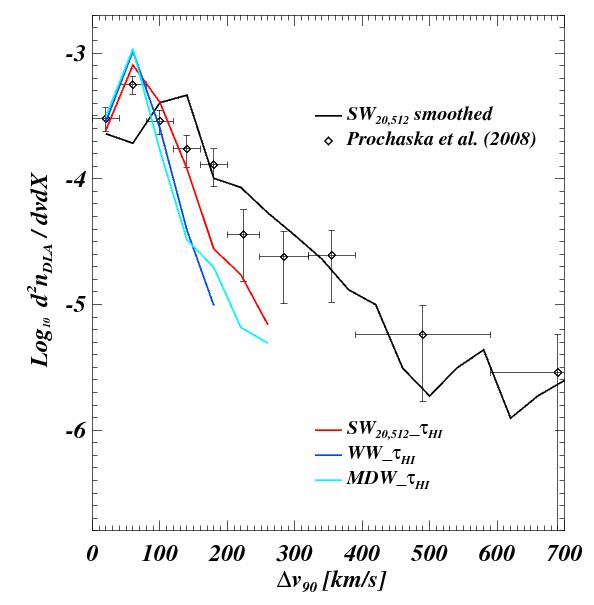}
\caption{{\it Left Panel}: DLAs velocity width distribution in runs
  SW$_{\rm 20,512}$ (dotted red line), WW (dot-dashed blue line), MDW
  (triple-dot-dashed cyan line) and SW$_{\rm 20,512}$ $smoothed$
  (solid black line) at $z=3$. {\it Right Panel}: DLAs velocity width
  distribution in runs SW$_{\rm 20,512}\_\tau_{\rm HI}$ (solid red
  line), WW$\_\tau_{\rm HI}$ (solid blue line), MDW$\_\tau_{\rm HI}$
  (solid cyan line) and SW$_{\rm 20,512}$ $smoothed$ (solid black
  line) at $z=3$. In both the panels the overplotted black diamonds
  show the observational data of \citet{prochaska08}.}
\label{fig:DLAveldistr}
\end{figure*}

In Figure \ref{fig:los} we compare three LOSs along the center of mass
(CM) of the second most massive halo in the SW, WW and MDW
simulations, at redshift $z=3$. We interpolate physical quantities
along the LOS using the SPH kernel of each particle.  The SW is
represented by the continuous black line, the WW is represented by the
dashed blue line and the MDW by the dotted red line. From top to
bottom we show the gas density, neutral hydrogen density, total
metallicity, temperature and peculiar velocity. The $x-$axis is in
km/s and represents at a given redshift the size of the box in real
space.  Looking at the first two top panels it is visible a peak both
in the gas and in the neutral hydrogen densities: this corresponds to
the CM of the halo. All simulations have more or less the same
structure for the peak in density but in the MDW case (red dotted
line) it is a bit more extended than the other two. The metallicity
panel clearly shows the effect of the different winds implementations
and velocities. The SW run produces a distribution of metallicity more
extended than MDW and WW while the WW distribution is narrower than
the MDW one. This confirms the fact that for massive haloes MDW is
less efficient than SW in extracting metals and distributing them in
the surrounding IGM (because of the high velocity dispersion of
massive haloes that results in a small loading factor for the wind
particles even if the corresponding wind velocities are high). The
temperature panel presents the same trends as the metallicity panel:
the high temperature peak is larger for SW with respect to MDW and WW,
demonstrating that the hot SW gas can be spread out to large
distances.  Quite interesting in this panel is the drop in temperature
at the center of the halo where the gas density is high and the
cooling efficient. The bottom panel shows the peculiar velocity field.
One can clearly see the effect of the expanding wind: a strong
discontinuity in the peculiar velocity produced by the shock, with a
negative peak (shell coming toward the observer) and a positive one
(shell expanding along the LOS in the opposite direction). Again the
SW have the larger distribution and WW the narrower. MDW have larger
peaks (in absolute values) than SW but not as extended, that are
produced by a fast, but less mass-loaded, wind. We note that the
peculiar velocity fields account for the observed wind velocities in
local starburst galaxies and that these velocities do depend on the
wind model.

\subsection{Velocity width distribution}
\label{sec:veldistr}
In this Section compare observational data and simulation predictions
on the observed velocity width distribution for low-ionization
species. We use the quantity $\Delta v_{\rm 90}$ defined as the extent
in velocity (redshift space) that embraces 90\% of the total
integrated optical depth as in \citet{prochaskawolfe97}. The procedure
to recover this statistics is the following: for every simulation we
take one thousands spectra extracted along the CMs of the most massive
haloes in the box. Then for every spectrum we compute the absorption
profile of the ion considered (SiII, $\lambda$ 1526.707 \AA) and
finally from this we determine the values of $\Delta v_{\rm 90}$
associated with that particular LOS (i.e with that particular DLA).
To deal with the fact that SiII occurs in self-shielded regions we use
the ISM multiphase model in the same way as we did for the
HI. Following Eqs. (1) and (2), we assign to each particle in the LOS
a mass in SiII:
\begin{eqnarray}
m_{\rm SiII}\,\,\,\,\, =&f_{\rm SiII}\, m_{\rm Si}\;\;\; & (\rho <
\rho_{\rm th})\\ m_{\rm SiII}\,\,\,\,\, =&f_{\rm c}\, m_{\rm Si}\;\;\;
&(\rho \geq \rho_{\rm th})\;,
\end{eqnarray}
where $m_{\rm Si}$ is the mass in Si of the particle (determined self
consistently inside the code), $f_{\rm SiII}$ is the SiII fraction as
determined by {\small CLOUDY} on the base of the temperature and the
density of the particle, $f_{\rm c}$ and $\rho_{\rm th}$ are,
respectively, the fraction of mass in cold clouds and the star
formation threshold (see Section 2).

It is well known that the observed distribution of velocity widths has
a median too large to be accomodated within standard cold dark matter
cosmology \citep{prochaska08, pontzen08}. In the left panel of Figure
\ref{fig:DLAveldistr} we show the DLA velocity width distributions at
redshift $z=3$ for our reference runs SW$_{\rm 20,512}$ (dotted red
line), WW (dot-dashed blue line), MDW (triple-dot-dashed cyan line)
along with observational data of \citet{prochaska08}. The latter
consist of a set of 113 measurements of metallicity, redshift and
$\Delta v_{\rm 90}$ taken with HIRES, ESI and UVES/VLT spectrographs
for DLAs in the redshift range $1.5<z<4.6$. We consider this sample as
a whole since it does not show any redshift evolution
\citep{pontzen08} and we compare with our $z=3$ outputs.

It is clear that our reference runs dramatically fail to reproduce
observational data even to a larger extent than found
by\citet{pontzen08}. In fact SW$_{\rm 20,512}$, WW and MDW
overpredicts the number of small velocity systems and at the same time
they are not able to produce systems with velocity width greater than
100 km/s. As we expect, MDW does a better job than the other two runs,
since it is more efficient in expelling metals, but it is yet far away
from fitting data. Checking one by one a large number of extracted
spectra we find that for a given LOS there is a significant number of
small regions in the optical depth, all of them with sizes that are
too small in redshift space. Basically our simulations, regardless to
the particular wind implementation, spread around the haloes small
clumps of enriched materials, that are in a wind phase, and are not
able to enrich uniformly the surrounding IGM. This of course also
results in the overproduction of small velocity systems seen in the
left panel of Figure \ref{fig:DLAveldistr}.

We checked that even using a simulation with higher resolution like
SW$_{\rm 10,448}$, which in principle should start enriching the IGM
earlier, we are not able to reproduce observational data. To
  overcome the problem we follow two other post-processing criteria
  with the goal of finding out some guidelines to improve our future
  simulations: SW$_{\rm 20,512} smoothed$ shown in both the panels of
  Figure \ref{fig:DLAveldistr} as the black solid line and the
  ``$\tau_{\rm HI}$'' series plotted in the right panel of Figure
  \ref{fig:DLAveldistr}.

For the $\tau_{\rm HI}$ series we take the original SW$_{\rm 20,512}$,
WW and MDW runs and then we follow \citet{pontzen08} assuming that
SiII is perfectly coupled to HI so that for solar metallicity $M_{\rm
  X}/M_{\rm H} = 0.0133$ and $n({\rm SiII})/n({\rm HI})=n({\rm
  Si})/n({\rm H})=3.47\times10^{-5}$. In this way SiII is spread more
efficiently around and inside the haloes. Moreover here we sum, for a
given LOS, all the different $\Delta v_{\rm 90}$ contribution due to
the small metal clumps trying to mimic a more uniform enrichment. The
results are shown in the right panel of Figure \ref{fig:DLAveldistr}:
SW$_{\rm 20,512}\_\tau_{\rm HI}$ is the solid red line, WW$\_\tau_{\rm
  HI}$ the solid blue line and MDW$\_\tau_{\rm HI}$ the solid cyan
line. SW$_{\rm 20,512}\_\tau_{\rm HI}$ works better than the other two
mostly because it has a greater number of bigger haloes where the
$\Delta v_{\rm 90}$ is correspondingly higher (see below), but as one
can see this post-processing procedure is not yet enough to match the
observations. We find a trend similar to what \citet{pontzen08} showed
in their Figure 9: all the different $\tau_{\rm HI}$ series reproduce
the small velocity tail of the distribution but fail to produce
systems with velocity width larger than $\sim$300 km/s.

Instead, for SW$_{\rm 20,512} smoothed$ we extract simulated spectra
after having increased the smoothing length associated to each
particle along the LOS and having recomputed the
metallicity. Basically we set the new smoothing length to 500 $h^{-1}$
comoving kpc and the result is the SW$_{\rm 20,512} smoothed$ curve
(black solid). With this prescription we can fit very well
observational data both at high and small velocities. Of course
spreading the metals a-posteriori over a scale of hundred $h^{-1}$kpc
is not self consistent but in this way we just want to see which
phenomenological prescription can be used in order to fit the data.

Either a pre-enrichment of the IGM at higher redshift that could
pollute the gas particles in a more uniform way
\citep[e.g.][]{tornatore07b} or some missing physical ingredient such
as small scale turbulence that could be effective in mixing metals in
a more efficient way at galactic scales
\citep[e.g.][]{scanna&bruggen08,iapichinoetal08}, could help in
reproducing the observed values. From this point of view a smoothing
scale of 500 $h^{-1}$kpc roughly corresponds to a velocity (in redshift
space) of $\sim$50 km/s, values that are consistent with those
presented in \citet{scanna&bruggen08}.

\begin{figure}
\includegraphics[width=8cm]{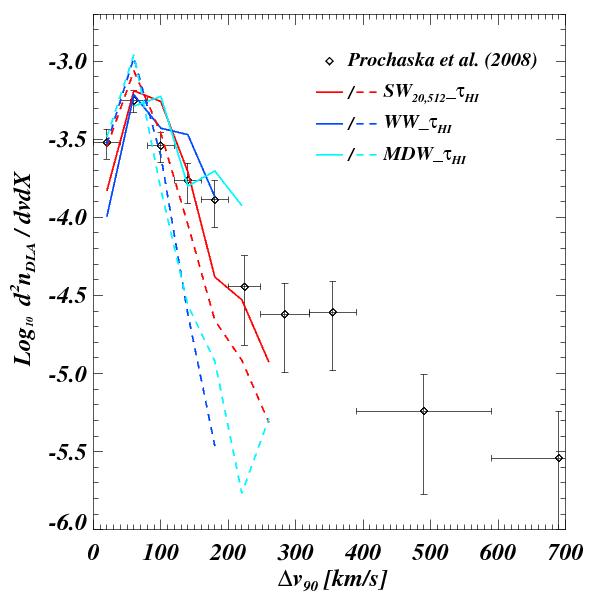}
\caption{Contribution to the DLAs velocity width distribution from
  haloes of different mass in runs SW$_{\rm 20,512}\_\tau_{\rm HI}$
  (solid + dashed red lines), WW$\_\tau_{\rm HI}$ (solid + dashed blue
  lines), MDW$\_\tau_{\rm HI}$ (solid + dashed cyan lines) at
  $z=3$. For the different runs, dashed lines refer to haloes with
  mass lower than $10^{10.5}h^{-1}M_{\rm \odot}$ and solid lines
  refers to haloes with mass greater than $10^{10.5}h^{-1}M_{\rm
    \odot}$. The overplotted black diamonds show the observational
  data of \citet{prochaska08}.}
\label{fig:PDFvm}
\end{figure}

Finally, in Figure \ref{fig:PDFvm} we test the contribution to the
DLAs velocity width distribution from haloes of different mass at
$z=3$. We use the three models that fit best the $\Delta v_{\rm 90}$
statistic, the $\tau_{\rm HI}$ series: SW$_{\rm 20,512}\_\tau_{\rm
  HI}$ (solid + dashed red lines), WW$\_\tau_{\rm HI}$ (solid + dashed
blue lines), MDW$\_\tau_{\rm HI}$ (solid + dashed cyan lines). For the
different runs, dashed lines refers to haloes with mass lower than
$10^{10.5}h^{-1}M_{\rm \odot}$ and solid lines refers to haloes with
mass larger than $10^{10.5}h^{-1}M_{\rm \odot}$. As we expect, massive
haloes produce much more systems with great velocity widths, while the
less massive haloes produce more smaller velocity systems. 

\begin{figure*}
\includegraphics[width=8.5cm]{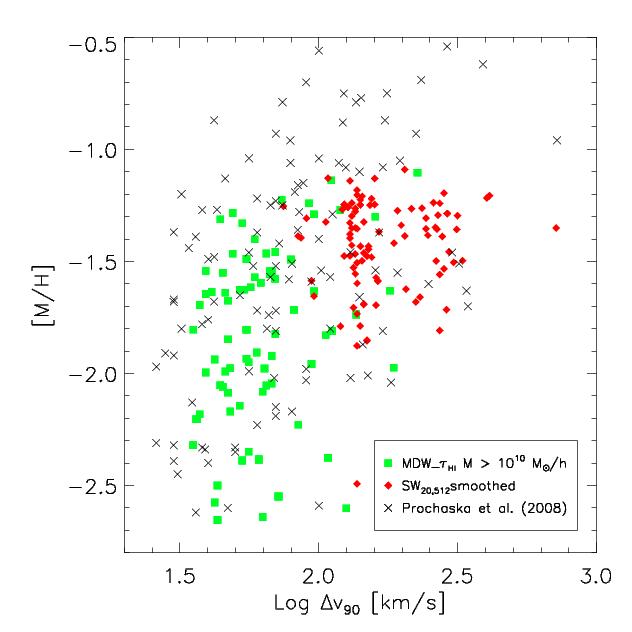}
\includegraphics[width=8.5cm]{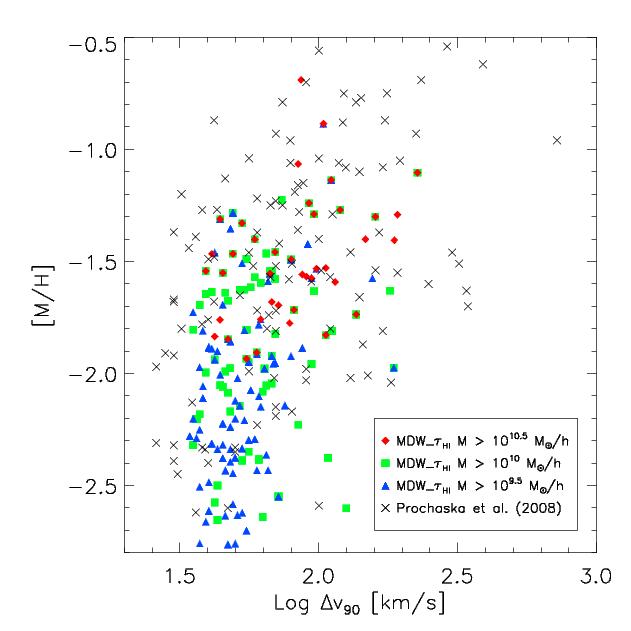}
\caption{{\it Left Panel}: relationship between metallicity and SiII
  velocity width of individual sightlines through haloes in
  MDW\_$\tau_{\rm HI}$ (green solid squares) and SW$_{\rm 20,512} smoothed$
  (red solid diamonds) at $z=3$, compared to the observational data by
  \citet{prochaska08} (black crosses). {\it Right Panel}: relationship
  between metallicity and SiII velocity width of individual sightlines
  through haloes with masses $M > 10^{10.5}h^{-1}M_{\rm \odot}$ (red solid
  diamonds), $M > 10^{10}h^{-1}M_{\rm \odot}$ (green solid squares) and $M >
  10^{9.5}h^{-1}M_{\rm \odot}$ (blue solid triangles) in MDW\_$\tau_{\rm HI}$ run at
  $z=3$, compared to the observational data by \citet{prochaska08}
  (black crosses).}
\label{fig:met_dvel_tau}
\end{figure*}

\subsection{Correlation between Metallicity and Velocity Widths}

Observational data by \citet{ledoux06} and \citet{prochaska08}
confirmed in the last years the existence of a positive correlation
between low-ion velocity width and the metallicity of DLAs along a
given sightline. We showed that our reference runs fail to reproduce
the DLAs velocity width distribution so we use the two post-processing
criteria presented in the previous Section (MDW$\_\tau_{\rm HI}$ and
SW$_{\rm 20,512} smoothed$) to test metallicity-velocity widths
correlation. In our simulations we compute the metallicities of DLAs
by taking the average metallicity value for the given LOS in a region
of $\pm$100 km/s in redshift space, centered at the location of the CM
of each halo hosting a DLA.

In the left panel of Figure \ref{fig:met_dvel_tau} we plot
metallicities vs. velocity widths for 100 haloes randomly selected in
a subsample of haloes with mass $M > 10^{10} h^{-1}M_{\rm \odot}$ in
the MDW$\_\tau_{\rm HI}$ (green solid squares) and SW$_{\rm 20,512}
smoothed$ (red solid diamonds), along with observational data by
\citet{prochaska08} (black crosses). Green squares (MDW$\_\tau_{\rm
  HI}$) show a weak correlation while correlation is negligible for
the red diamond (SW$_{\rm 20,512} smoothed$). SW$_{\rm 20,512}
smoothed$ have in average both metallicities and velocity widths
higher than MDW$\_\tau_{\rm HI}$ and this is not surprising due to the
procedure of smoothing metallicity over a large scale. For this plot
we use only haloes with masses above $M > 10^{10} h^{-1}M_{\rm \odot}$
because we noted that for the most massive haloes the correlation is
stronger.  In the right panel of Figure \ref{fig:met_dvel_tau} we
clarify this point using the MDW$\_\tau_{\rm HI}$ run. Red diamonds
refers to all the haloes with mass $M > 10^{10.5} h^{-1}M_{\rm
  \odot}$, green squares are the same green points of the left panel
(a subsample of haloes with mass $M > 10^{10} h^{-1}M_{\rm \odot}$),
while blue triangles derive from a subsample of haloes with mass $M >
10^{9.5} h^{-1}M_{\rm \odot}$. It is evident that the correlation in
this case is somewhat weaker. Instead, using a subsample of haloes at
larger masses, we increase the value of $\Delta v_{\rm 90}$, since
faster winds are produced by more massive haloes, that usually are
more metal rich at least when compared with smaller mass haloes.
Thereby, if we restrict ourselves to haloes above
$10^{10.5}h^{-1}$\msun the agreement is reasonably good, even though
smaller mass haloes, as we saw in the previous sections, contribute to
all of the other DLA statistical properties. We also note that in
general the mean metallicities recovered from our simulations are
somewhat smaller than the observed ones.

\section{Discussion and Conclusions}
In this paper we investigated the properties of DLAs in
high-resolution hydrodynamical cosmological simulations.  In
particular, we focussed on the role of feedback in the form of
galactic winds to quantify their impact on the neutral hydrogen and
metal ion species content both on the global IGM and on the IGM around
putative DLA sites.  The analysis was made by comparing with the
recent works of \citet{nagamine07} and \citet{pontzen08}, but with
important differences with respect to these previous analyses: $i)$
higher resolution simulations (by a factor $\sim 10$ in mass) than
\citet{nagamine04,nagamine07}; $ii)$ statistics performed in a
cosmological setting, like in \cite{nagamine04,nagamine07}, but
differently from \citet{pontzen08}, who relied mainly on simulations
of single objects; $iii)$ inclusion of metal cooling and an accurate
chemo-dynamical code that follows self consistently the IGM enrichment
\citep{T07}; $iv)$ inclusion of different wind prescriptions (energy
driven and momentum driven); $v)$ modification of the stellar IMFs and
the linear dark matter power spectrum in the initial conditions to
explore the parameter space further.

The conclusions that we draw from our analysis can be summarized as
follows.
\begin{itemize}
\item[-] The different feedback prescriptions that we explored (WW,
  weak energy-driven winds of 100 km/s; SW, strong energy-driven winds
  of 600 km/s; MDW, momentum driven winds) give distinct predictions
  for the gas distribution in metallicity-temperature and density.  SW
  are effective in heating the gas particles at temperatures of $10^5$
  K to a larger extent than MDW and WW. Also the metallicity-density
  distribution is different: WW and MDW show correlations in
  high-density regions (denser regions being more metal rich), while
  SW is more efficient in polluting the low-density IGM (Figures 1 and
  2).
\item[-] WW, SW and MDW have also an impact on global quantities of
  the simulations such as the cosmic star formation rates and the
  neutral hydrogen content. As for the latter, MDW and SW show
  agreement with data while in the WW the HI is clearly overproduced
  since the feedback is not as efficient. Also, the star formation
  rate for the WW run is higher than for the SW one (see Figures 3 and
  14).
\item[-] The evolution with redshift of CIV and OVI (two of the most
  common ion species observed in QSO absorption spectra) is similar in
  shape between the different runs but the normalization differ by up
  to a factor 3 (see Figure 4).
\item[-] Focussing on the properties of the haloes that could host
  DLAs we find that haloes between $10^9-10^{10}h^{-1}M_{\rm \odot}$
  at $z=3$ have similar SFR for the different wind models of about
  $0.01-0.1$ \msun/yr, with large scatter, while more massive haloes
  have larger SFRs and the trend with mass depends on the specific
  wind model (Figure 7). The metallicities of the haloes as a function
  of mass is correlated for haloes of masses $>10^{9.5}h^{-1}M_{\rm
    \odot}$ in the WW and MDW runs, while no correlation or possibly a
  weak anti-correlation is seen for the SW model. For less massive
  haloes below $10^{9.5}h^{-1}$ $M_{\rm \odot}$ the scatter in
  metallicity is huge and span several orders of magnitude (Figure
  8).
\item[-] The cross-sections inferred from the different runs are in
  overall agreement with the results of \citet{nagamine04,nagamine07},
  although the values for shape and normalization are somewhat smaller
  for similar runs. This is likely to be due to the different
  prescriptions of the wind implementation and the metal cooling. The
  DLA incidence rate is in good agreement with the observed one by
  \citet{prochaska05} from SDSS data for MDW and SW, while is larger
  for WW. The incidence rate results clearly show that all the haloes
  whose masses are above $10^9h^{-1}M_{\rm \odot}$ contribute to the
  cross-section (Figures 10 and 11).
\item[-] The column density distribution function at $z=3$ is in rough
  agreement with the data points for all the models (for MDW the
  agreement is better), while at lower redshifts there are larger
  differences and SW (WW) underpredicts (overpredicts) the number of
  DLAs. The WW run shows the largest discrepancies especially for the
  high column density DLAs, a feature that was already found by
  \citet{nagamine04}.  The contribution of haloes of masses between
  $10^9-10^{10}h^{-1}M_{\rm \odot}$ is particularly relevant for DLAs
  below $N_{\rm HI} \approx 10^{20.8}$ cm$^{-2}$, another hint that
  less massive haloes contribute to reproduce DLAs statistical
  properties (Figures 12 and 13).
\item[-] Qualitatively, physical quantities interpolated along LOS
  that pierce the haloes show different behaviours (particularly in
  temperature and metallicity) in different models. Also the peculiar
  velocity fields look different (Figure 15) and the gas velocity
  gradient is in agreement with the observed wind velocities.
\item[-] The distribution of velocity widhts of low ioniziation
  species (in our case SiII) reproduces the observed one
  \citep{prochaska08} only for velocity widths below 100 km/s, while
  larger values are dramatically under-reproduced by our
  simulations. A slightly better agreement is obtained when SiII is
  supposed to faithfully trace the HI distribution \citep[as done
    by][]{pontzen08}, but even in this case the observed distribution
  is underestimated. We can fit the observed velocity widths well only
  if we empirically smooth the metallicity of each gas particle over a
  region of 500 comoving $h^{-1}$kpc (Figure 16). We also find a trend
  of $\Delta v_{90}$ with the halo mass that goes in the expected
  direction: more massive haloes produce the larger values for $\Delta
  v_{90}$ (Figure 17).
\item[-] The metallicity-velocity width correlations is broadly
  reproduced by the two models discussed above, even if the
  metallicity values are somewhat lower than the observed ones. If we
  split the contribution to the correlation by taking into account
  different ranges of masses, we find that there is good agreement if
  we restrict ourselves to the most massive haloes with
  $M>10^{10.5}h^{-1}M_{\rm \odot}$ (Figure 18). This is in agreement
  with \citet{barnes08}, but in disagreement with other statistical
  properties of DLAs, that in order to be reproduced need less massive
  haloes as well.
\end{itemize}

Overall, we succeeded in reproducing most of the observed properties
of DLAs in particular the column density distribution function and the
incidence rate. It appears that the best agreement is given either by
the SW or the MDW implementation, while the WW does not seem to fit
the data.  For the metal distributions, as traced by the velocity
widths, the agreement is not good. This could be the hint that there
are some pieces of physics missing in our treatment. Possible
ingredients that are not considered here and that could help in easing
the discrepancy are: $i)$ radiative transfer effects in DLAs that have
been either neglected so far \citep[e.g.][]{nagamine04} or
approximately modelled \citep[e.g.][]{pontzen08}: even if their impact
seems to be not dramatic at least using our effective model
description of the ISM, since the cold clouds are assumed to be fully
self-shielded and the ambient medium to be optically thin, they could
be important for large column density systems, especially when
affecting the molecular hydrogen in star forming regions
\citep{gnedin08}; $ii)$ small-scale turbulence and its impact on the
metal diffusion at large scales, which has been recently investigated
by \citet{scanna&bruggen08} and that can be effective in smoothing the
metal distribution around DLAs; $iii)$ a pre-enrichment of the IGM,
possibly produced by PopIII stars, at higher redshift than those
considered here that could result in a smoother metallicity for the
particles that are in a wind phase \citep[e.g.][]{tornatore07b}.  We
stress that all these effects are not considered in our simulations
and their accurate description could be important to match the
remaining still unexplained DLA properties.

\section*{Acknowledgments}
The authors thank F. Biondi, A. Saro, E. Spitoni, A. Bignamini and
G. Vladilo for many helpful discussions and acknowledge the anonymous
referee for the helpful report.  Numerical computations were done on
the COSMOS (SGI Altix 3700) supercomputer at DAMTP and at High
Performance Computer Cluster (HPCF) in Cambridge (UK) and at CINECA
(Italy).  COSMOS is a UK-CCC facility which is supported by HEFCE,
PPARC and Silicon Graphics/Cray Research.  The CINECA (``Centro
Interuniversitario del Nord Est per il Calcolo Elettronico'') CPU time
has been assigned thanks to an INAF-CINECA grant. This work has been
partially supported by the INFN-PD51 grant, an ASI-AAE Theory grant
and a PRIN-MIUR.

\bibliographystyle{mn2e}
\bibliography{tescari_aph}

\label{lastpage}
\end{document}